\def\squareSummable {\mathcal{L}^2({\mathbb{R}})}
\def\integrald{{\rm d}}
\def\complex{{j}}
\def\timeSpacingVariable{\tau_0}
\def\frequencySpacingVariable{\nu_0}
\def\timeSpacing{T}
\def\frequencySpacing{F}
\def\symbols[#1][#2][#3]{{X_{#1#2}^{#3}}}
\def\estimatedSymbols[#1][#2][#3]{{\tilde{X}_{#1#2}^{#3}}}
\def\transmittedSignal[#1][#2]{s^{#2}(#1)}
\def\aSignal[#1][#2]{s_{#2}(#1)}
\def\receivedSignal{r(\timeSymbol)}
\def\frequencyIndexTX{l}
\def\timeIndexTX{n}
\def\timeSymbol{t}
\def\timeShift{\tau}
\def\frequencyShift{\nu}
\def\frequencyIndexRX{k}
\def\timeIndexRX{m}
\def\numberOfSubcarrier{N}
\def\activity[#1]{{Q_{#1}^{\timeIndexTX}}}

\def\translatedModulatedFilterTXDiscrete[#1][#2]{g_{#1#2}}
\def\translatedModulatedFilterDiscrete[#1][#2]{\gamma_{#1#2}}
\def\translatedModulatedFilterCDiscrete[#1][#2]{\gamma^*_{#1#2}}

\def\translatedModulatedFilterTX[#1][#2][#3]{g_{#1#2}^{\epsilon}(#3)}
\def\translatedModulatedFilterTXAggressor[#1][#2][#3]{g_{#1#2}^{\aggressorindex}(#3)}
\def\translatedModulatedFilter[#1][#2][#3]{\gamma^{\epsilon}_{#1#2}(#3)}
\def\translatedModulatedFilterC[#1][#2][#3]{{\gamma^\epsilon}^*_{#1#2}(#3)}
\def\pulseShaping[#1]{p\left(#1\right)}

\def\channelProcess[#1]{H_{#1}(\timeShift,\frequencyShift)}
\def\channelGlobal[#1]{A_{\timeIndexTX\frequencyIndexTX\timeIndexRX\frequencyIndexRX}^{#1}}
\def\channelGlobalSelf[#1]{A_{\timeIndexRX\frequencyIndexRX\timeIndexRX\frequencyIndexRX}^{#1}}
\def\truncation{K}

\def\ppp{\Phi}

\def\prototypeFilterTXAggressor[#1]{g^\aggressorindex\left(#1\right)}
\def\prototypeFilterTX[#1]{g^\epsilon\left(#1\right)}
\def\prototypeFilterRX[#1]{\gamma^{\epsilon}\left(#1\right)}
\def\prototypeFilterRXc[#1]{{\gamma^{\epsilon}}^*\left(#1\right)}

\def\channelProcessTime[#1]{h_{#1}(\timeShift,\timeSymbol)}

\def\gaussianRollOff{\rho}

\def\aggressorPathlossConstant{a_{\rm }}
\def\aggressorPathlossSlope{b_{\rm }}

\def\aggressorVictimPathloss{{L}_{\rm m}}

\def\carrierFrequency{f_{\rm c}}

\def\fc{f_{\rm c}}

\def\numberOfPath{L}
\def\channelPathIndex{\ell}

\def\pathGain{\varrho_\channelPathIndex(\timeSymbol)}

\def\noisePower{\sigma_{\rm noise}^{2}}
\def\pathlosscompensationParameter{\beta}

\def\timingMisalignmentTime{{\Delta t_{{\it \aggressorindex}}}}
\def\intentionalCFO{{\Delta f_{{\it \aggressorindex}}}}
\def\iCFOindex{r}
\def\iCFOset{\Omega}
\def\iCFO[#1]{\psi_{#1}}
\def\iCFOprobability[#1]{p_{#1}}

\def\victimDistanceOwnUser{d_{{\epsilon}}}

\def\victimnoise{W_{\frequencyIndexRX}}
\def\victimnoiseTime[#1]{w\left(#1\right)}
\def\aggressordistance[#1]{r_{{\rm #1}\aggressorindex}}
\def\aggressorDistanceVariable{\nu}

\def\aggressorDensity{\lambda}

\def\aggressorMinimumDistance{r_{\rm min}}
\def\aggressorindex{i}

\def\aggressorRxPower{P_{{\epsilon}}}
\def\aggressordistanceOwnUser{d_{\aggressorindex}}
\def\aggressordistanceOwnUserVariable{u}

\def\aggressorInterferencePower{P_{{\aggressorindex}}}

\def\interferenceFunction{I_{\rm total}}
\def\interferenceFunctionAggressor{I_\aggressorindex}
\def\interferenceFunctionAggressorAll{I_{\rm other}}
\def\interferenceFunctionSelf{I_{\rm self}}

\def\translatedModulatedVectorFilterTX[#1][#2]{p_{#1,#2}}
\def\translatedModulatedVectorFilter[#1][#2]{\acute{p}_{#1,#2}}
\def\translatedModulatedVectorFilterC[#1][#2]{\acute{p}_{#1,#2}^*}

\def\frame{u_{\timeIndexRX\frequencyIndexRX}(\timeSymbol)}

\def\constellationIndex{q}
\def\modulationOrder{M}
\def\modulationConstant{c_\constellationIndex}

\def\SINR{\rm SINR}
\def\EbN0{ {E_{\rm b}}/{N_0}}
\def\H0{H_0(\frequencyIndexRX)}

\def\interferenceGain[#1]{{G_{#1}}}
\def\meanInterferenceGain[#1][#2]{\sigma_{{#1}}^2(#2)}
\def\meanSelfInterferenceGain{\sigma_{\rm self}^2}
\def\lambdaInterferenceGain[#1]{\lambda_{{#1}}}
\def\rollOff{\alpha}
\def\figureDistance{-3mm}

\def\expectedPowerOperator[#1]{
\mathbb{E}\left[{{\left|{#1}\right|}^2}\right]
}

\def\expectedOperator[#1][#2]{
\mathbb{E}_{#2}\left[{#1}\right]
}

\def\LaplaceOperator[#1][#2]{
{\mathcal{L}}_{#1}\left( \displaystyle #2 \right)
}

\def\diracOperator[#1]{
{\delta}\left({#1}\right)
}

\def\kroneckerOperator[#1]{
{\delta}_{#1}
}

\def\pdfOperator[#1]{
{f_{{#1}}}\left({#1}\right)
}

\def\probabilityOperator[#1]{
{\rm Prob}\left\{{#1}\right\}
}

\def\absOperator[#1]{
\left|{#1}\right|
}

\def\erfcOperator[#1]{
{\rm erfc}\left({#1}\right)
}

\def\gammaOperator[#1]{
{\rm \Gamma}\left({#1}\right)
}

\def\lambdaOperator[#1][#2]{
{\lambda}\left({#1}\right)
}

\def\COperator[#1][#2]{
{C}\left({#1}\right)
}

\def\berOperator[#1]{
{\rm BER}\left({#1}\right)
}

\def\positiveComplexRotation[#1][#2]{{\rm e}^{\complex 2\pi\frac{#1#2}{\numberOfSubcarrier}}}

\def\negativeComplexRotation[#1][#2]{{\rm e}^{-\complex 2\pi\frac{#1#2}{\numberOfSubcarrier}}}

\documentclass[journal]{IEEEtran}
\usepackage{stfloats}
\usepackage{xcolor}
\usepackage{amsfonts}
\usepackage{acronym}
\usepackage{cite}
\usepackage{amsmath,amssymb}
\usepackage[dvips]{graphicx}
\usepackage[caption=false,font=footnotesize]{subfig}
\usepackage[geometry]{ifsym}
\usepackage{flushend}

\begin{document}

%
\title{\huge Partially Overlapping Tones for Uncoordinated Networks}

\author{\authorblockN{Alphan \c{S}ahin$^1$, Erdem Bala$^2$,  {\.I}smail G\"{u}ven\c{c}$^3$, Rui Yang$^2$, and H\"{u}seyin Arslan$^1$\\}
\authorblockA{$^1$Department of Electrical Engineering, University of South Florida, Tampa, FL, 33620}\\
\authorblockA{$^2$InterDigital Communications Inc.,
Huntington Quadrangle, Melville, NY 11747}\\
\authorblockA{$^3$Department of Electrical and Computer Engineering, Florida International University, Miami, FL, 33174}\\
Email: {\tt alphan@mail.usf.edu}, {\tt {erdem.bala}@interdigital.com},  {\tt iguvenc@fiu.edu}, {\tt rui.yang@interdigital.com}, {\tt arslan@usf.edu}
}


\maketitle

\begin{abstract}
In an uncoordinated network, the link performance between the devices might
degrade significantly due to the interference from other
links in the network sharing the same spectrum. As a solution, in this study, the concept of {\em partially overlapping tones (POT)}  is introduced. The interference energy observed at the victim receiver is mitigated by partially overlapping the individual subcarriers via an intentional carrier frequency offset between the links. Also,
it is shown that while orthogonal transformations at the receiver cannot mitigate the other-user interference without losing spectral efficiency, non-orthogonal transformations are able to mitigate the other-user interference without any spectral efficiency loss at the expense of self-interference. Using spatial Poisson point process, a tractable bit error rate analysis is provided to demonstrate potential benefits emerging from POT.  
\end{abstract}


\IEEEpeerreviewmaketitle
\acrodef{PMF}[PMF]{probability mass function}
\acrodef{HetNets}[HetNets]{heterogeneous networks}
\acrodef{MGF}[MGF]{moment generation function}
\acrodef{PDF}[PDF]{probability density function}
\acrodef{PSWF}[PSWF]{prolate spheroidal wave function}
\acrodef{OFDM}[OFDM]{Orthogonal frequency division multiplexing}
\acrodef{NOFDM}[NOFDM]{non-orthogonal frequency division multiplexing}
\acrodef{OFDMA}[OFDMA]{orthogonal frequency division multiple accessing}
\acrodef{BFDM}[BFDM]{biorthogonal frequency division multiplexing}
\acrodef{DFT}[DFT]{discrete Fourier transformation}
\acrodef{IDFT}[IDFT]{inverse discrete Fourier transformation}
\acrodef{IFFT}[IFFT]{inverse fast Fourier transformation}
\acrodef{FBMC}[FBMC]{filter bank multicarrier}
\acrodef{CP}[CP]{cyclic prefix}
\acrodef{PAPR}[PAPR]{peak-to-average-power ratio}
\acrodef{QAM}[QAM]{quadrature amplitude modulation}
\acrodef{OQAM}[OQAM]{offset quadrature amplitude modulation}
\acrodef{FFT}[FFT]{fast Fourier transformation}
\acrodef{RRC}[RRC]{root-raised-cosine}
\acrodef{RC}[RC]{raised-cosine}
\acrodef{TC}[TC]{tapered-cosine}
\acrodef{CFO}[CFO]{carrier frequency offset}
\acrodef{SIR}[SIR]{signal-to-interference ratio}
\acrodef{SNR}[SNR]{signal-to-noise ratio}
\acrodef{SINR}[SINR]{signal-to-interference-plus-noise ratio}
\acrodef{ICI}[ICI]{inter-carrier interference}
\acrodef{ACI}[ACI]{adjacent-channel interference}
\acrodef{CCI}[CCI]{co-channel interference}
\acrodef{ISI}[ISI]{inter-symbol interference}
\acrodef{PPN}[PPN]{poly phase network}
\acrodef{WSSUS}[WSSUS]{wide-sense stationary uncorrelated scattering}
\acrodef{SEM}[SEM]{spectral emission mask}
\acrodef{BWA}[BWA]{broadband wireless access}
\acrodef{PSD}[PSD]{power spectral densitie}
\acrodef{MIMO}[MIMO]{multiple-input multiple-output}
\acrodef{DSL}[DSL]{digital subscriber lines}
\acrodef{OFDP}[OFDP]{optimal finite duration pulse}
\acrodef{FMT}[FMT]{filtered multitone}
\acrodef{SMT}[SMT]{staggered multitone}
\acrodef{CMT}[CMT]{cosine-modulated multitone}
\acrodef{IMT}[IMT]{isotropic multitone}
\acrodef{IOTA}[IOTA]{isotropic orthogonal transform algorithm}
\acrodef{RMS}[RMS]{root mean square}
\acrodef{MMSE}[MMSE]{minimum mean square error}
\acrodef{MLD}[MLD]{maximum likelihood detection}
\acrodef{STBC}[STBC]{space time block coding}
\acrodef{FTD}[FTD]{fractional time delay}
\acrodef{IAM}[IAM]{interference approximation method}
\acrodef{PN}[PN]{phase noise}
\acrodef{RF}[RF]{radio-frequency}
\acrodef{CPM}[CPM]{continuous phase modulation}
\acrodef{ADC}[ADC]{analog-to-digital converter}
\acrodef{PA}[PA]{power amplifier}
\acrodef{CDF}[CDF]{cumulative distribution function}
\acrodef{SC-FDMA}[SC-FDMA]{single carrier frequency division multiple accessing}
\acrodef{FPGA}[FPGA]{field-programmable gate array}
\acrodef{STTC}[STTC]{space-time trellis coding}
\acrodef{BER}[BER]{bit error rate}
\acrodef{ZP}[ZP]{zero padding}
\acrodef{FDE}[FDE]{frequency domain equalization}
\acrodef{TDE}[TDE]{time domain equalization}
\acrodef{LS}[LS]{least square}
\acrodef{SC-FDE}[SC-FDE]{single carrier frequency domain equalization}
\acrodef{FB-S-FBMC}[FB-S-FBMC]{filter bank spread FBMC}
\acrodef{LTE}[LTE]{Long Term Evolution}
\acrodef{AWGN}[AWGN]{additive white Gaussian noise}
\acrodef{RD}[RD]{random demodulator}
\acrodef{FTN}[FTN]{faster-than-Nyquist}
\acrodef{PRS}[PRS]{partial-response signaling}
\acrodef{MAP}[MAP]{maximum a posteriori}
\acrodef{DFE}[DFE]{decision feedback equalizer}
\acrodef{FIR}[FIR]{finite impulse response}
\acrodef{LMS}[LMS]{least-mean-square}
\acrodef{RLS}[RLS]{recursive-least-squares}
\acrodef{PHYDYAS}[PHYDYAS]{Physical layer for dynamic access}
\acrodef{MLSE}[MLSE]{maximum likelihood sequence estimator}
\acrodef{RP}[RP]{reception point}
\acrodef{TP}[TP]{transmission point}
\acrodef{PPP}[PPP]{Poisson point processe}
\acrodef{D2D}[D2D]{device-to-device}
\acrodef{TO}[TO]{timing offset}
\acrodef{I-factor}[I-factor]{interference factor}
\acrodef{PSK}[PSK]{phase shift keying}
\acrodef{UWB}[UWB]{ultra wide band}
\acrodef{MAC}[MAC]{media access control}
\acrodef{SIC}[SIC]{successive interference cancellation}
\acrodef{POT}[POT]{partially overlapping tones}
\acrodef{iid}[i.i.d.]{independent and identically distributed}

\section{Introduction}

Traditional broadband wireless networks have been strained with emerging demands such as being always-connected to the network and very high throughput to satisfy data-hungry applications such as real-time video. Satisfaction of these demands constitutes the main driving force for \ac{HetNets} in which multiple tiers with varying coverage co-exist over the same network. In \ac{HetNets}, interference among the tiers or the devices might dominate the noise and create interference-limited networks. The interference issues become prominent especially when dense and unplanned deployments such as \ac{D2D} communications are taken into account.  Considering this issue, a new technique in which certain features of the waveform itself are used to mitigate the interference is proposed.

A waveform, which is one of the core elements determining the characteristics of a communication system, describes the formation of associated resources in signal space \cite{shannon_1998,sahin2013Survey}. Robustness of the transmitted signal to dispersion in the transmission medium, channel access, and hardware complexity are just few features affected by the selected waveform. Hence, waveform design should be able to address the requirements specified by the system. When the performance of the network is limited by noise, main consideration for the waveform design can naturally be on individual link properties such as reducing the interference created by the time and frequency dispersion of the channel \cite{Kozek_1998}.  However, interference created by other users is many times a major factor limiting the performance of a network and as such the impact of the other-user interference might be more significant compared to the interference due to the channel dispersion.  {\em Conventionally}, the interference between the devices are elaborated with the approaches which question the amount of the interference power at the receiver location without including the impact of the waveform itself. 
Most of the solutions devised to address the interference problem rely either on \ac{MAC} based coordination or interference cancellation. For example, interference coordination mechanisms with proper scheduling and resource allocation aim to minimize the interference power
\cite{Lopez-Perez_2011}. In physical layer, methods like interference cancellation \cite{Andrews_SIC}, multiuser detection \cite{verdu1998multiuser}, and interference alignment \cite{Cadambe_2008} handle the other-user interference by exploiting the difference between desired and interfering signal strengths, codes, and multipath channel.


As opposed to the conventional solutions, in this paper, a new concept based on utilizing the time-frequency characteristics of waveforms to reduce the other-user interference is proposed. The main contributions of this paper are:
\begin{itemize}
	\item We introduce the concept of \ac{POT} in which it is allowed for subcarriers allocated to interfering links to partially overlap. The overlap is achieved by introducing an intentional CFO between the links and its amount is controlled by appropriately designing the time-frequency utilization of the waveforms.
	\item It is shown that with orthogonal waveforms, there is a tradeoff between other-user interference and spectral efficiency. Mitigation of the other-user interference can be achieved at the expense of a loss in spectral efficiency.
	\item It is further shown that with non-orthogonal waveforms, there is a tradeoff between other-user interference and self-interference. Mitigation of the other-user interference can be achieved at the expense of increased self-interference  while spectral efficiency remains unchanged.
	\item A tractable \ac{BER} analysis for an uncoordinated network deployment is provided. The analysis allows to understand the system performance for various network densities and waveform designs.
\end{itemize}

The rest of paper is organized as follows: Related work is discussed in Section \ref{sec:relatedWorks}. The system model including the physical layer parameters is provided in Section \ref{sec:system_model} while the concept of POT for orthogonal and non-orthogonal waveform structures is introduced in Section \ref{sec:waveforms}. Then, \ac{BER} analysis is provided in Section \ref{sec:BERana} and numerical results evaluating the performance of the proposed approach are provided in Section \ref{sec:numericalResults}. Finally, the paper is concluded in Section \ref{sec:conclusion}.

\section{Related Work}
\label{sec:relatedWorks}

The concept of overlapping wireless channels exists within the several 802.11 families (e.g. Wi-Fi systems).  However, the simultaneous access to the channels is usually avoided due to interference. The utilization of overlapping channels to improve throughput has been investigated in several papers \cite{Mishra_2006, Ding_2008, Cui_2011,ileri_2011,duarte_2012,Fu_2013}. In \cite{Mishra_2006}, it is emphasized that the channel separation between the two pairs of Wi-Fi nodes can be interpreted as the physical separation between the nodes. Therefore, if partially overlapping channels are used carefully, it can provide greater spatial re-use.  
These papers consider the total spectrum utilization of the transmission, and do not show the impact of the partial overlapping on  {\em individual subcarriers}.  
To the best of our knowledge, detailed time-frequency analysis on the interference due to the partially overlapping  pulse shapes  is not  available in the literature.

Some of the challenging aspects of the other-user interference are its asynchronous nature and its statistical characterization, which depend on the deployment model and waveform structure utilized in the network. \ac{OFDM} is a well-investigated multicarrier scheme in case of asynchronous interference, e.g., femtocell-macrocell coexistence  \cite{Vikram_CM_2008,Sahin_Opp_2009,sahin_icc2011}. By providing some timing offset between the tiers intentionally, the different types of the interference, i.e. \ac{ICI} and \ac{ISI}, is converted into each other in \cite{sahin_icc2011}. Yet, the total other-user interference is kept constant. 
A theoretical \ac{BER} analysis investigating   \ac{ISI} versus \ac{ICI} trade-offs in \ac{OFDM}
 downlink is provided  in \cite{Hamdi_2009}. 
In \cite{34_Medjahdi_TWC_2011}, \ac{BER} degradation due to the adjacent channel interference is investigated by emphasizing superiority of \ac{FBMC} based cellular systems over an \ac{OFDM} based approach. Although these investigations provide useful intuitions on the performance degradation, the analyses are performed for idealistic assumptions, such as grid-based cell deployment and uniform user density. In \cite{sahin_2012_user}, it is emphasized that even if the geographical user density is uniform, the distance of the users linked to the corresponding serving points might not be uniform due to the irregular base station deployment and shadowing characteristics. 
In \cite{Win_2009,andrews_2011},
homogeneous \acp{PPP} are considered to model the deployment of the base stations. This approach, which is pessimistic compared to highly idealized grid-based models and real deployment scenarios, yields a tractable tool which exploits the stochastic geometry. 
In the following studies, e.g., \cite{novlan_2012} and \cite{Dhillon_2012}, analytical models for uplink and $K$-tier heterogeneous networks are provided using \acp{PPP}.

Investigation on the impact of \acp{PPP} on physical layer is limited, but available. For example, coexistence between \ac{UWB} and
narrow band systems is investigated using \acp{PPP} and impact of pulse shape is emphasized for aggregate network emission \cite{Win_2009}. In \cite{Pinto_2010}, error rate analyses are provided for \ac{QAM} and \ac{PSK} modulations using \acp{PPP}, excluding the impact of waveforms.

\section{System Model}

\label{sec:system_model}

\begin{figure}[!t]
\centering
{\includegraphics[width =2.3in]{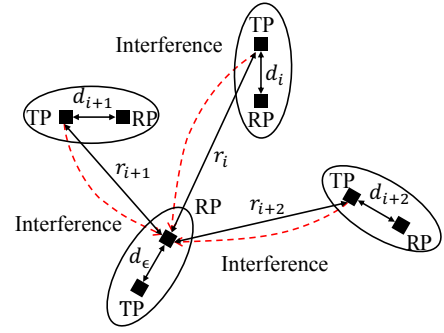}
}
\caption{Illustration of interference in an uncoordinated network.}
\label{fig:model}
\end{figure} 
Consider an uncoordinated network where \acp{TP} and their corresponding \acp{RP} are distributed in an area as a realization of homogeneous 2-D \ac{PPP} of $\ppp$ with the intensity $\aggressorDensity$ as in \figurename~\ref{fig:model}. 
Interfering \acp{TP} and the \ac{RP} investigated are called as {\em aggressors} and {\em victim}, respectively. 
Without any loss of generality, victim is located at the origin of the polar coordinates (0,0). The distance between the $\aggressorindex$th aggressor and the victim is given as $\aggressordistance[]$. Minimum distance between the aggressors and the victim is set to $\aggressorMinimumDistance$. While the distance between \ac{RP} and its associated \ac{TP} for $\aggressorindex$th aggressor link is denoted by $\aggressordistanceOwnUser$, the same distance is expressed by 
$\victimDistanceOwnUser$ for the desired link for the victim. Also,
it is assumed that aggressors are farther away than $\victimDistanceOwnUser$, i.e., $ \aggressordistance[] > \aggressorMinimumDistance \ge \victimDistanceOwnUser$, which is widely considered for the interference analyses based on \acp{PPP} \cite{lin_2013}. 

In the following subsections, signal model for transmission and reception based on multicarrier schemes and channel model that includes large and small scale effects are given for further discussions on \ac{POT}.

\subsection{Signal Model for Transmission}
The transmitted signal from the desired \ac{TP} and the transmitted signals from the $\aggressorindex$th aggressor can be expressed as
\begin{align}
\transmittedSignal[\timeSymbol][\epsilon] = &\sum_{\timeIndexTX = -\infty}^{\infty}\sum_{\frequencyIndexTX =0}^{\numberOfSubcarrier-1} \symbols[\timeIndexTX][\frequencyIndexTX][\epsilon]  \translatedModulatedFilterTX[\timeIndexTX][\frequencyIndexTX][\timeSymbol]~,
\label{Eq:signal_tx_Epsilon}
\end{align}
and
\begin{align}
\transmittedSignal[\timeSymbol][\aggressorindex] = &\sum_{\timeIndexTX = -\infty}^{\infty}\sum_{\frequencyIndexTX =0}^{\numberOfSubcarrier-1} \symbols[\timeIndexTX][\frequencyIndexTX][\aggressorindex]  \translatedModulatedFilterTXAggressor[\timeIndexTX][\frequencyIndexTX][\timeSymbol]~,
\label{Eq:signal_tx_MS}
\end{align}
respectively, where ${\symbols[\timeIndexTX][\frequencyIndexTX][\epsilon]}$ and ${\symbols[\timeIndexTX][\frequencyIndexTX][\aggressorindex]}$ are the information symbols which are \ac{iid} with zero mean on the
$\frequencyIndexTX$th subcarrier and $\timeIndexTX$th  symbol,  
$\numberOfSubcarrier$ is the number of subcarriers, and
$\translatedModulatedFilterTX[\timeIndexTX][\frequencyIndexTX][\timeSymbol]$ and $\translatedModulatedFilterTXAggressor[\timeIndexTX][\frequencyIndexTX][\timeSymbol]$ are the {\em synthesis functions} which map information symbols into time-frequency plane based on a rectangular lattice as
\begin{align}
\translatedModulatedFilterTX[\timeIndexTX][\frequencyIndexTX][\timeSymbol]&=\prototypeFilterTX[\timeSymbol - \timeIndexTX \timeSpacingVariable]e^{j2\pi\frequencyIndexTX \frequencySpacingVariable\timeSymbol}
\label{Eq:lattice}
\end{align}
and
\begin{align}
\translatedModulatedFilterTXAggressor[\timeIndexTX][\frequencyIndexTX][\timeSymbol]&=\prototypeFilterTXAggressor[\timeSymbol - \timeIndexTX \timeSpacingVariable]e^{j2\pi\frequencyIndexTX \frequencySpacingVariable\timeSymbol}~.
\label{Eq:latticeAggressor}
\end{align}
The family of functions in \eqref{Eq:lattice} and \eqref{Eq:latticeAggressor} are often referred to as {\em Gabor frame} or  {\em Weyl-Heisenberg frame}, where $\prototypeFilterTX[\timeSymbol]$ and $\prototypeFilterTXAggressor[\timeSymbol]$ are the prototype filters employed at the transmitters, $\frequencySpacingVariable$ is the subcarrier spacing and $\timeSpacingVariable$ is the symbol spacing \cite{Gabor_1946,97_Sondergaard_Thesis_2007}. For the sake of  notation simplicity, $\frequencySpacingVariable$ and $\timeSpacingVariable$ are given in units of $\frequencySpacing$ and $\timeSpacing$, respectively (e.g., $\frequencySpacingVariable=1.2\times\frequencySpacing$ and $\timeSpacingVariable=1.3\times\timeSpacing$), where $\frequencySpacing = 1/\timeSpacing$ and $\frequencySpacing$ is a number based on the design.
Without loss of generality, the energy of $\prototypeFilterTX[\timeSymbol]$ and the energy of $\prototypeFilterTXAggressor[\timeSymbol]$ are normalized as
\begin{align}
\lVert\prototypeFilterTX[\timeSymbol]\rVert^2_{\squareSummable}=\lVert\prototypeFilterTXAggressor[\timeSymbol]\rVert^2_{\squareSummable}&=\int_{-\infty}^{\infty}\absOperator[{\prototypeFilterTX[\timeSymbol ]}]^2\integrald\timeSymbol=1~,
\label{Eq:energy}
\end{align}
where $\squareSummable$ denotes the square-integrable function space over $\mathbb{R}$ and $\lVert \cdot \rVert$ is the $\mathcal{L}^2$-norm of function. 


\subsection{Large Scale Impacts}
Considering various path loss models depending on the environment, the path loss  is characterized by $\aggressorVictimPathloss(\cdot)  = \aggressorPathlossConstant+\aggressorPathlossSlope\log_{10}(\cdot)$
where the path loss parameters $\aggressorPathlossConstant$ and $\aggressorPathlossSlope$ are scalars and the argument is the distance in meters. The received interference power from the $\aggressorindex$th aggressor  and the desired signal power at victim location per subcarrier are denoted by $\aggressorInterferencePower$ and $\aggressorRxPower$, respectively.
 Impact of shadowing is not considered in this study. Main reason for this issue is to give insights on the \ac{POT} rather than introducing extra complexity for the system model. However, using the methodologies proposed for the moment generation function of the summations of lognormal distributed lognormal variables  \cite{Tellambura_2010} and \cite{Mehta_2007}, it is possible to include the impact of shadowing on the investigation.  

For the link transmission, open loop fractional power control is applied and some amount of the path loss, i.e., $\pathlosscompensationParameter (\aggressorPathlossConstant+\aggressorPathlossSlope\log_{10}(\cdot))
$, is compensated, where $\pathlosscompensationParameter \in [0,1]$ is the path loss compensation parameter. Note that \ac{TP} might transmit with the maximum transmit power in some cases. However, since link distances considered are small, the possibility of transmission at maximum power is excluded.

\subsection{Small Scale Impacts}
Time-varying multipath channel is taken into account between all \acp{RP} and \acp{TP}. Channel impulse response is characterized by
$
\channelProcessTime[] = \sum_{\channelPathIndex = 0}^{\numberOfPath - 1} \pathGain\diracOperator[\timeShift - \timeShift_\channelPathIndex]
$ where $\numberOfPath$ denotes the total number of multipaths, $\channelPathIndex$ is the path index, and $\timeShift_\channelPathIndex$ is the delay of the $\channelPathIndex$th path. It is assumed that the path gains, $\pathGain$, are independent and identically distributed variables and the signals experience Rayleigh fading, which is a common model for interference analysis. Also, the expected channel power is considered as $\sum_{\channelPathIndex = 0}^{\numberOfPath - 1}\expectedPowerOperator[{\pathGain}] = 1$. For the sake of notation, the channel between $\aggressorindex$th interfering \ac{TP} and the victim {RP} and the channel between desired \acp{TP} and the victim \ac{RP} are expressed as $\channelProcessTime[\aggressorindex]$ and $\channelProcessTime[\epsilon]$, respectively.

\subsection{Synchronization}
As discussed in \cite{Marey_2007} and \cite{Morelli_2008}, synchronization to the received signal in the presence of interference might be challenging, especially at low \ac{SINR}s. However, the impairments like timing offset and \ac{CFO} are often related to the preamble structure rather than the data portion of the frame. Therefore, perfect synchronization at the pair of interest is assumed. Besides, timing misalignment between the aggressor's signals and synchronization point of the victim  is taken into account. 
The timing misalignment of $\aggressorindex$th aggressor signal with respect to the synchronization point of the victim \ac{RP} is denoted by
$\timingMisalignmentTime$ and its distribution  $\pdfOperator[\timingMisalignmentTime]$ is assumed as uniform between 0 and $\timeSpacingVariable$.
Besides, intentional \ac{CFO} between $\aggressorindex$th aggressor and the victim \ac{RP} is given by $\intentionalCFO$ in order to generate \ac{POT} which is discussed in Section \ref{sec:waveforms}.
The impact of \ac{CFO} due to the hardware mismatches between the aggressor's signals and desired signal is ignored. This is because of the fact that the impact of CFO due to the hardware mismatches is relatively smaller than $\intentionalCFO$ for \ac{POT}. For example, when carrier spacing is set to 15 kHz and CFO is 500 Hz,  normalized \ac{CFO} becomes 0.033 ($500 $ Hz / $15$ kHz). However, the amount of normalized $\intentionalCFO$ for \ac{POT}, throughout the study, is at least 0.5, which is significantly larger than \ac{CFO} due to the hardware error.

\subsection{Signal Model for Reception}
Considering all interfering \acp{TP}, and assuming a \ac{WSSUS} channel model \cite{85_bello1963characterization}, the received signal at the victim is obtained as
\begin{align}
&\receivedSignal= \underbrace{\displaystyle\sqrt{\aggressorRxPower} \int_{\timeShift}\int_{\frequencyShift}\channelProcess[\epsilon] 
 \transmittedSignal[\timeSymbol -\timeShift][\epsilon]e^{j2\pi \frequencyShift\timeSymbol}\integrald\frequencyShift\integrald\timeShift }_{\rm Desired~signal} \nonumber \\ &\textrm{+}\underbrace{\displaystyle\sum_{\aggressorindex \in\ppp}\sqrt{\aggressorInterferencePower}\int_{\timeShift}\int_{\frequencyShift}\channelProcess[\aggressorindex] 
 \transmittedSignal[\timeSymbol +\timingMisalignmentTime-\timeShift][\aggressorindex]e^{j2\pi \frequencyShift\timeSymbol}\integrald\frequencyShift\integrald\timeShift}_{\rm Interfering~signals}
\textrm{+}\underbrace{
\victimnoiseTime[\timeSymbol]}_{\rm Noise}
\label{Eq:receivedSignalDoppler}
\end{align}
where $\channelProcess[\epsilon]$ and $\channelProcess[\aggressorindex]$ are the Fourier transformations of  $\channelProcessTime[\epsilon]$ and $\channelProcessTime[\aggressorindex]$, respectively, and $\victimnoiseTime[\timeSymbol]$ is the \ac{AWGN} with zero mean and variance $\noisePower$. In order to get the information symbol on the
$\frequencyIndexRX$th subcarrier and $\timeIndexRX$th symbol, 
the received signal is correlated by the {\em analysis function} where
\begin{align}
\translatedModulatedFilter[\timeIndexRX][\frequencyIndexRX][\timeSymbol]&= \prototypeFilterRX[\timeSymbol - \timeIndexRX \timeSpacingVariable]e^{j2\pi\frequencyIndexRX \frequencySpacingVariable\timeSymbol}~.
\label{Eq:latticeR}
\end{align}
Then, the output of the correlator is sampled with the sampling period to obtain the received symbol as 
\begin{align}
\estimatedSymbols[\timeIndexRX][\frequencyIndexRX][\epsilon]&=\langle \receivedSignal, \translatedModulatedFilter[\timeIndexRX][\frequencyIndexRX][\timeSymbol]\rangle 
\triangleq\int_{\timeSymbol} \receivedSignal \translatedModulatedFilterC[\timeIndexRX][\frequencyIndexRX][\timeSymbol] d\timeSymbol\nonumber\\
&=\underbrace{\sqrt{\aggressorRxPower} \symbols[\timeIndexRX][\frequencyIndexRX][\epsilon]\channelGlobalSelf[\epsilon]}_{\rm desired ~part}
+\underbrace{\sqrt{\aggressorRxPower}\sum_{\substack{ \timeIndexTX =-\truncation+1 \\ \timeIndexTX \neq\timeIndexRX}}^{\truncation-1} \sum_{
\substack{\frequencyIndexTX =0\\ \frequencyIndexTX \neq \frequencyIndexRX }
}^{\numberOfSubcarrier-1}\symbols[\timeIndexTX][\frequencyIndexTX][\epsilon]\channelGlobal[\epsilon]}_{\rm self\textrm{-}interference~part}
 \nonumber\\&+\underbrace{\sum_{\aggressorindex \in\ppp}\sqrt{\aggressorInterferencePower}\sum_{\substack{ \timeIndexTX =-\truncation+1  }}^{\truncation-1} \sum_{
\substack{\frequencyIndexTX =0 }
}^{\numberOfSubcarrier-1}\symbols[\timeIndexTX][\frequencyIndexTX][\aggressorindex]\channelGlobal[\aggressorindex]}_{\rm other\textrm{-}user~interference}+\underbrace{\victimnoise}_{\rm noise}~.
\label{Eq:receivedSymbol}
\end{align}
In \eqref{Eq:receivedSymbol},
\begin{align}
&\channelGlobal[\epsilon] \textrm{=}  \int_{\timeShift}\int_{\frequencyShift}\channelProcess[\epsilon]\int_{\timeSymbol} \translatedModulatedFilterTX[\timeIndexTX][\frequencyIndexTX][\timeSymbol-\timeShift]
\translatedModulatedFilterC[\timeIndexRX][\frequencyIndexRX][\timeSymbol]
e^{j2\pi \frequencyShift\timeSymbol}{\integrald\timeSymbol}{\integrald\frequencyShift}{\integrald\timeShift}~,
\label{Eq:receivedInterferenceSymbol}\\&
\channelGlobal[\aggressorindex] \textrm{=} \int_{\timeShift}\int_{\frequencyShift}\channelProcess[\aggressorindex]\int_{\timeSymbol} \translatedModulatedFilterTXAggressor[\timeIndexTX][\frequencyIndexTX][\timeSymbol-\timingMisalignmentTime-\timeShift]
e^{j2\pi\intentionalCFO(\timeSymbol-\timingMisalignmentTime-\timeShift)}\nonumber\\&
~~~~~~~~~~~~~~~~~~~~~~~~~~~~~~~~~\times\translatedModulatedFilterC[\timeIndexRX][\frequencyIndexRX][\timeSymbol]
e^{j2\pi \frequencyShift\timeSymbol}{\integrald\timeSymbol}{\integrald\frequencyShift}{\integrald\timeShift}~,
\label{Eq:receivedInformationSymbol}
\end{align}
and they show the correlation between the symbols ($\timeIndexTX,\frequencyIndexTX $) and ($\timeIndexRX,\frequencyIndexRX $) including the dispersion due the channel.
As it is seen in \eqref{Eq:receivedSymbol}, while other-user interference is caused by aggressor links, self-interference can occur due to the time-varying multipath channel, hardware impairments, or non-Nyquist filter utilization.
Considering \eqref{Eq:receivedSymbol}, \ac{SINR} can be expressed as
\begin{align}
&{\SINR}\nonumber=\\&\frac{\overbrace{\absOperator[{\channelGlobalSelf[\epsilon]}]^2}^{\interferenceGain[\epsilon]}}{\displaystyle \underbrace{
\underbrace{\sum_{\substack{ \timeIndexTX =-\truncation+1 \\ \timeIndexTX \neq\timeIndexRX}}^{\truncation-1} \sum_{
\substack{\frequencyIndexTX =0\\ \frequencyIndexTX \neq \frequencyIndexRX }
}^{\numberOfSubcarrier-1}\absOperator[{\channelGlobal[\epsilon]}]^2}_{\interferenceFunctionSelf}\textrm{+}
\underbrace{\sum_{\aggressorindex\in\ppp }\underbrace{\frac{{\aggressorInterferencePower}  }{{\aggressorRxPower}} \underbrace{\sum_{\substack{ \timeIndexTX =-\truncation+1  }}^{\truncation-1}\sum_{
\substack{\frequencyIndexTX =0 }
}^{\numberOfSubcarrier-1}\absOperator[{\channelGlobal[\aggressorindex]}]^2}_{\interferenceGain[ \aggressorindex]}}_{\interferenceFunctionAggressor}}_{ \interferenceFunctionAggressorAll}}_{\interferenceFunction}\textrm{+}   \frac{\noisePower}{{\aggressorRxPower}}}~,
\label{Eq:SINR} 
\end{align}
 where $\truncation$ is the filter length in terms of symbol spacing, $\interferenceFunction$ is the total interference, $\interferenceFunctionSelf$ and $\interferenceFunctionAggressorAll$ are the self-interference and other-user interference, respectively, $\interferenceFunctionAggressor$ is the interference due to $\aggressorindex$th aggressor, $\interferenceGain[\epsilon]$ and $\interferenceGain[\aggressorindex]$ are the interference gains including fading and filter characteristics, and
\begin{align}
\frac{{\aggressorInterferencePower}  }{{\aggressorRxPower}} = 
{\victimDistanceOwnUser}^{\frac{\aggressorPathlossSlope-\pathlosscompensationParameter\aggressorPathlossSlope}{10}}
\aggressordistanceOwnUser^{\frac{\pathlosscompensationParameter\aggressorPathlossSlope}{10}}{\aggressordistance[]}^{\frac{-\aggressorPathlossSlope}{10}}~. \label{Eq:COperator}
\end{align}
Note that $\truncation$ is related to the representation of the filter in time domain. As long as $\truncation$ is selected properly, the filter truncation has a minor impact on self-interference compared to the interference due to the time-varying multi-path channel or hardware impairments at the \ac{RP} and/or \ac{TP}.
While $\interferenceGain[\epsilon]$ is a random variable with unit mean exponential distribution because of the Rayleigh fading \cite{Hamdi_2009,34_Medjahdi_TWC_2011}, $\interferenceGain[\aggressorindex]$ can be characterized for a given $\timingMisalignmentTime$ and $\intentionalCFO$ by exponential distribution where its mean is given by 
\begin{align}
\meanInterferenceGain[{\aggressorindex}][{\timingMisalignmentTime,\intentionalCFO}]= {{ \sum_{\timeIndexTX =-\truncation+1 }^{\truncation-1}\sum_{
\frequencyIndexTX =0 }^{\numberOfSubcarrier-1}\absOperator[{\langle \translatedModulatedFilterTXAggressor[\timeIndexTX][\frequencyIndexTX][\timeSymbol-\timingMisalignmentTime]
e^{j2\pi\intentionalCFO\timeSymbol}
,\prototypeFilterRX[\timeSymbol]
}\rangle]^2}}~,\label{Eq:other}
\end{align}
Conventionally, $\meanInterferenceGain[{\aggressorindex}][{\timingMisalignmentTime,\intentionalCFO}]$ is  considered as $1$ for link-level analyses \cite{andrews_2011}, similar to the mean of $\interferenceGain[\epsilon]$. However, expressing it as in \eqref{Eq:other} gives flexibility to include the impact of transmit and receive filters and  calculate interference when an additional processing is performed to reduce other-user interference. Finally, $\interferenceFunctionSelf$ is also a random variable with exponential distribution where, considering the Rayleigh fading assumption \cite{34_Medjahdi_TWC_2011}, its mean is given by
\begin{align}
\meanSelfInterferenceGain=
 {{ \sum_{\substack{ \timeIndexTX =-\truncation+1 \\ \timeIndexTX \neq\timeIndexRX}  }^{\truncation-1}\sum_{\substack{\frequencyIndexTX =0\\ \frequencyIndexTX \neq \frequencyIndexRX }}^{\numberOfSubcarrier-1}\absOperator[{\langle \translatedModulatedFilterTX[\timeIndexTX][\frequencyIndexTX][\timeSymbol]
,\prototypeFilterRX[\timeSymbol]
}\rangle]^2}}
~.\label{Eq:self}
\end{align}
Essentially, calculations of both $\meanSelfInterferenceGain$ and $\meanInterferenceGain[{\aggressorindex}][{\timingMisalignmentTime,\intentionalCFO}]$  are  based on the projection operation onto receive filters, which can be derived via corresponding ambiguity functions \cite{sahin2013Survey}.

\section{Partially Overlapping Tones}
\label{sec:waveforms}

\begin{figure}[!t]
\centering
\subfloat[Fully overlapping tones. ]{\includegraphics[width =1.65in]{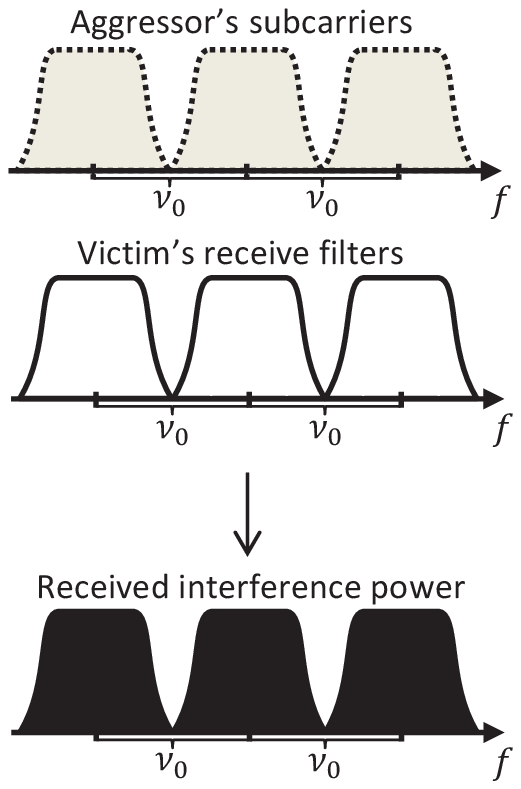}
\label{fig:full}}
\subfloat[Partially overlapping tones. ]{\includegraphics[width =1.65in]{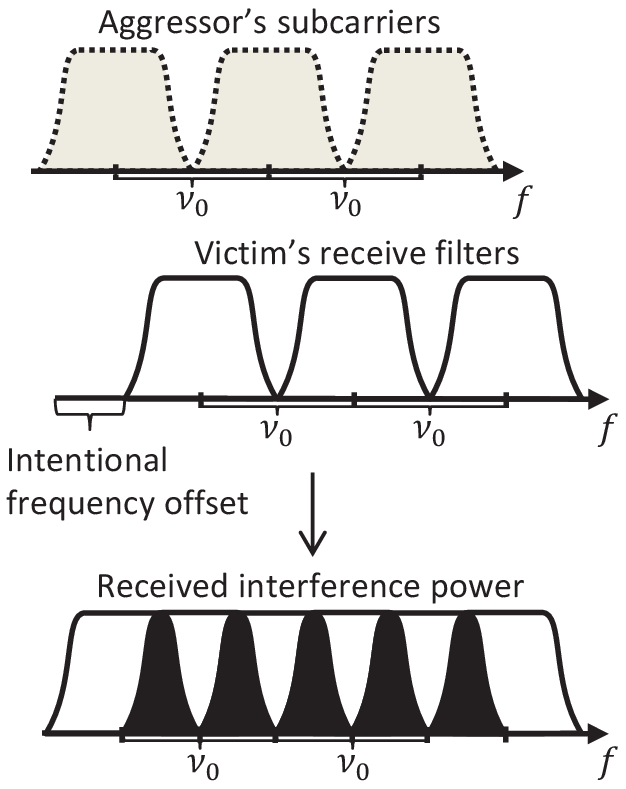}
\label{fig:partial}}
\caption{Illustrations for full overlapping and partial overlapping. While full-overlapping tones cause significant other-user interference, the main portion of the interference is mitigated by the receive filter with the concept of POT.}
\label{fig:overlappingTones}
\end{figure}

The main goal of the \ac{POT} is to mitigate other-user interference given in \eqref{Eq:other} by using the waveform structure. It relies on intentional \ac{CFO} between aggressor's Gabor system and victim's Gabor system. For example, while one of the links operates at carrier frequency $\fc$, the other link operates at $\fc+\frequencySpacingVariable/2$. By allowing this operation, instead of full-overlapping between the subcarriers of the links, \ac{POT} is obtained. 
 This approach also fits the asynchronous nature of other-user interference as it does not introduce any timing constraint between interfering signals. One can interpret the intentional \ac{CFO} as an alignment strategy in frequency domain. 

In \figurename~\ref{fig:overlappingTones}, a motivating example based on \ac{FMT}   is illustrated for \ac{POT}.
In \ac{FMT}, each subcarrier is generated via a band-limited filter \cite{16_cherubini2002filtered}. As opposed to the conventional understanding of \ac{OFDM}, the subcarriers are not overlapped in frequency domain. By providing additional guard bands, orthogonality between subcarriers is maintained. 
Note that these guard bands are also useful to provide immunity against self-interference due to the time-frequency impairments. 
In the provided example in  \figurename~\ref{fig:overlappingTones}, these guard bands are exploited further and they are used to mitigate the other-user interference. By applying an intentional \ac{CFO} between two different links, other-user interference mitigation is provided in an uncoordinated network.

\ac{POT} is fundamentally related to the utilization of the time-frequency plane by the waveform structure. Transmit filter, receiver filter, and density of symbols in time-frequency plane determine the available resource opportunities jointly for the other-user interference mitigation by using \ac{POT}, as exemplified in \figurename~\ref{fig:overlappingTones}. Besides,  further utilization of the waveform structure via non-orthogonal schemes along with \ac{POT} lead to a trade-off for  uncoordinated networks: {\em other-user interference versus self-interference}. This trade-off is desirable in an uncoordinated network as long as self-interference is handled via self-interference cancellation methods, e.g., equalization. In the following subsections, orthogonality of schemes is stressed in conjunction with \ac{POT}. \ac{POT} with orthogonal schemes and non-orthogonal schemes
are investigated theoretically along with numerical results
and their potential drawbacks.

\subsection{Partially Overlapping Tones with Orthogonal Schemes}
\label{subsec:tradeoff1}

 For orthogonal schemes, transmitter and receiver utilize the same prototype filter, i.e., $\translatedModulatedFilterTX[\timeIndexRX][\frequencyIndexRX][\timeSymbol]=\translatedModulatedFilter[\timeIndexRX][\frequencyIndexRX][\timeSymbol]$. In addition, inner products of the different basis functions derived from the prototype filter yield zero correlations, i.e., $\langle\translatedModulatedFilterTX[\timeIndexTX][\frequencyIndexTX][\timeSymbol],\translatedModulatedFilter[\timeIndexRX][\frequencyIndexRX][\timeSymbol]\rangle =\kroneckerOperator[\timeIndexTX\frequencyIndexTX\timeIndexRX\frequencyIndexRX]$.   Many fundamental schemes, e.g., \ac{OFDM}, \ac{FMT}, and \ac{FBMC}, rely on orthogonality.
In digital communication, orthogonality in a multicarrier scheme is generally perceived as a necessary condition. It simplifies the receiver algorithms significantly and provides optimum \ac{SNR} performance in \ac{AWGN} channels. Besides these features, orthogonal schemes have another fundamental property due to orthogonal basis functions at the receiver: 
the energy of a signal before the projection onto receive filters is equal to the energy after the projection onto receiver filters. This is typically expressed through the Plancherel formula\footnote{It corresponds to Parseval's theorem for Fourier series.} given by
\begin{align}
\lVert \aSignal[\timeSymbol][] \lVert^2 = \sum_{\timeIndexRX,\frequencyIndexRX} | \langle \aSignal[\timeSymbol][], \frame \rangle |^2~,
\label{Eq:orthogonal}
\end{align}
where $\aSignal[\timeSymbol][]$ is an arbitrary signal, and $\{\frame\}$ is a set of orthogonal basis functions. 
Assume that $\aSignal[\timeSymbol][]$ is the interfering signal. When an orthogonal transformation, e.g., \ac{DFT}, is applied to $\aSignal[\timeSymbol][]$ at the receiver, the total amount of the interference does not change after the transformation. This issue leads to an undesirable result: only way to {\it mitigate} the other-user interference is to discard some of subcarriers or to construct an {\em incomplete} Gabor system, i.e., $\timeSpacingVariable\frequencySpacingVariable>1$ \cite{Kozek_1998, sahin2013Survey}, which causes less spectrally efficient schemes.  In other words, \ac{POT} with orthogonal schemes would be beneficial only when some of subcarriers are not utilized or $\timeSpacingVariable\frequencySpacingVariable>1$. Indeed, norm-preserving feature of orthogonal transformations at the receivers explain {\em why orthogonal schemes  do not directly provide immunity against the other-user interference}.

\ac{POT} offers intentional \ac{CFO} between the different links based on the fact that timing synchronization between \acp{TP}  in an uncoordinated network is a challenging issue. However, the  intentional \ac{CFO} approach also introduces some constraints on the waveform structure. For example, orthogonal multicarrier schemes which provide non-overlapping subcarriers in frequency domain, e.g., \ac{FMT}, complies with the intentional \ac{CFO} approach introduced by \ac{POT}. However, \ac{POT} might not be as beneficial as in the case of \ac{FMT} to the schemes where the orthogonality is maintained strictly on certain localizations in the time-frequency plane, as in \ac{OFDM}. Considering this issue, analyses throughout the study are performed based on \ac{FMT}.

\begin{figure}[t]
\centering
\subfloat[ Impact of timing misalignment when root raised cosine filter is employed along with FMT.]{\includegraphics[width=3.5in]{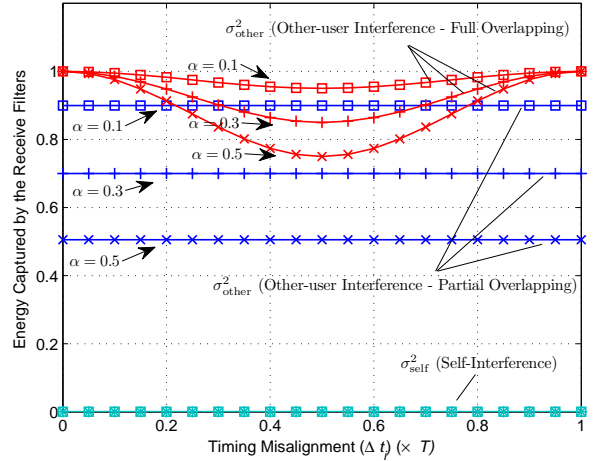}
\label{fig:fmt}}\\
\subfloat[ Trade-off between spectral efficiency and other-user interference.]
{\includegraphics[width=3.5in]{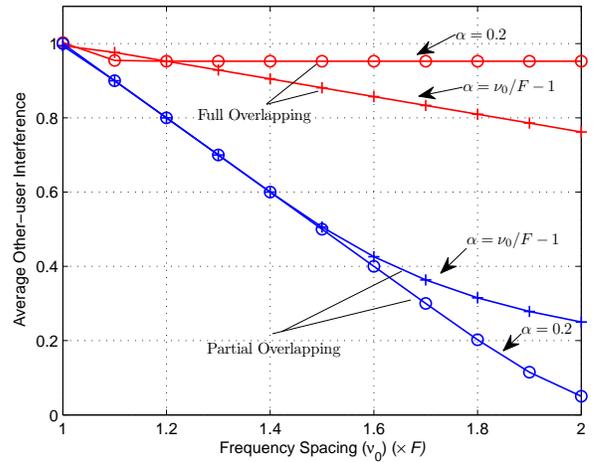}\label{fig:tradeOff_lossy}
}
\caption{Other-user interference mitigation without introducing self-interference, but loss in spectral efficiency.}
\label{fig:interference_lossy}
\end{figure}

In \figurename~\ref{fig:interference_lossy}, 
considering timing misalignment between one aggressor and the victim, $\timingMisalignmentTime$ is swept for one symbol period when  $\intentionalCFO=\frequencySpacingVariable/2$. \ac{FMT} is generated based on \ac{RRC} filter. Note that \ac{RRC} filter is a band-limited filter and the excess bandwidth of the \ac{RRC} filter is controlled via a roll-off factor of $\rollOff$, where $0\le\rollOff\le 1$. 
In \figurename~\ref{fig:interference_lossy}\subref{fig:fmt}, $\meanInterferenceGain[{\aggressorindex}][{\timingMisalignmentTime,\intentionalCFO}]$  is calculated numerically, based on \eqref{Eq:other}.
In case of full overlapping, $\meanInterferenceGain[{\aggressorindex}][{\timingMisalignmentTime,\intentionalCFO}]$ is mitigated maximally when $\timingMisalignmentTime = 0.5\times \timeSpacing$, $\timeSpacingVariable=\timeSpacing$, and $\frequencySpacingVariable=(1+\rollOff)\times\frequencySpacing$. This is because of  the reduction of the \ac{ICI} components  maximally due to the additional guard bands, when timing misalignment occurs. 
In case of partial overlapping, impact of $\timingMisalignmentTime$ is removed totally, and $\meanInterferenceGain[{\aggressorindex}][{\timingMisalignmentTime,\intentionalCFO}]$ is significantly reduced since the receive filters reject the main portion of the interference, depending on the utilized $\rollOff$. 
Assuming the aggressor interference has a uniform timing misalignment characteristics, trade-off between spectral efficiency and other-user interference is given for two different \ac{FMT} cases in \figurename~\ref{fig:interference_lossy}\subref{fig:tradeOff_lossy}.
When $\frequencySpacingVariable$ is set to $(1+\rollOff)\times\frequencySpacing$, $\meanInterferenceGain[{\aggressorindex}][{\timingMisalignmentTime,\intentionalCFO}]$ decreases for both full overlapping and partial overlapping due to the less \ac{ICI} components with the timing misalignment, as given in \figurename~\ref{fig:interference_lossy}\subref{fig:fmt}. When $\rollOff$ is fixed to $0.2$, other-user interference is mitigated more via partial overlapping, since this approach provides more gap in frequency for other-user interference mitigation.

Major concern of using \ac{POT} with orthogonal schemes might be having less spectral efficient transmission for the sake of other-user interference mitigation. However, as indicated before, it allows the devices interrupted by the interference to achieve a better \ac{BER} performance with a simple approach.

\subsection{Partially Overlapping Tones with Non-orthogonal Schemes}
\label{subsec:tradeoff2}

Similar to the orthogonal schemes, transmitter and receiver utilize the same prototype filters for non-orthogonal structures, i.e., $\translatedModulatedFilterTX[\timeIndexRX][\frequencyIndexRX][\timeSymbol]=\translatedModulatedFilter[\timeIndexRX][\frequencyIndexRX][\timeSymbol]$. However, inner products of the different basis functions do not yield zero correlations, i.e., $\langle\translatedModulatedFilterTX[\timeIndexTX][\frequencyIndexTX][\timeSymbol],\translatedModulatedFilter[\timeIndexRX][\frequencyIndexRX][\timeSymbol]\rangle \neq\kroneckerOperator[\timeIndexTX\frequencyIndexTX\timeIndexRX\frequencyIndexRX]$. For example, \ac{NOFDM} can be constructed by using the rectangular lattice of \ac{OFDM} with non-Nyquist transmit filters and receive filters, e.g., Gaussian functions. For non-orthogonal schemes, the utilized basis functions at the receiver also corresponds to a nonorthogonal transformations, i.e., $\langle\translatedModulatedFilter[\timeIndexTX][\frequencyIndexTX][\timeSymbol],\translatedModulatedFilter[\timeIndexRX][\frequencyIndexRX][\timeSymbol]\rangle \neq\kroneckerOperator[\timeIndexTX\frequencyIndexTX\timeIndexRX\frequencyIndexRX]$. In that case, the condition given in \eqref{Eq:orthogonal} is relaxed as 
\begin{align}
A\lVert \aSignal[\timeSymbol][] \lVert^2 \le \sum_{\timeIndexRX,\frequencyIndexRX} | \langle \aSignal[\timeSymbol][], \frame \rangle |^2 \le B\lVert \aSignal[\timeSymbol][]\lVert^2~,
\label{Eq:nonorthogonal}
\end{align}
where $\{\frame\}$ is a set of non-orthogonal elements, $A$ and $B$ are the lower bound and upper bound, respectively, and $0<A\le B< \infty$. Based on \eqref{Eq:nonorthogonal}, when a non-orthogonal transformation is applied at the receiver, the energy of $\aSignal[\timeSymbol][]$ does not have to be preserved after the transformation. In other words, the non-orthogonal transformations at the receivers are able to alter the amount of the observed interference energy. Hence, when \ac{POT} is taken into account with non-orthogonal schemes, it is possible to mitigate other-user interference even when $\timeSpacingVariable\frequencySpacingVariable=1$.


\begin{figure}[!t]
\centering
\subfloat[Less self-interference, but more other-user-interference.  ]{\includegraphics[width=3.3in]{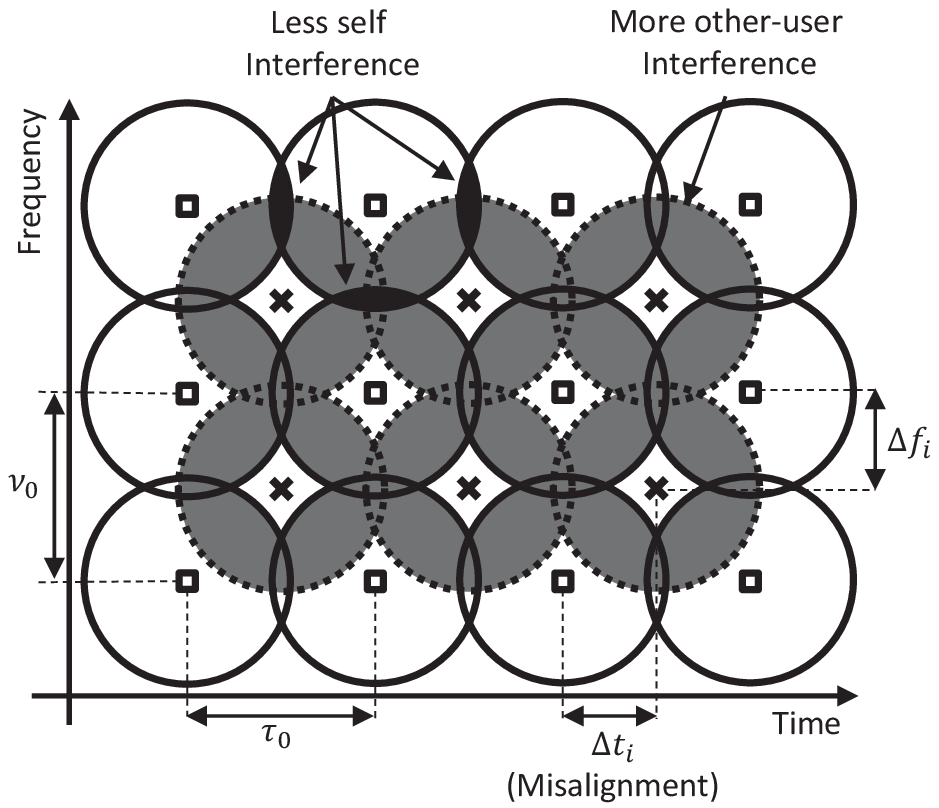}
\label{fig:i1}}\\
\subfloat[More self-interference, but less other-user-interference. ]{\includegraphics[width=3.3in]{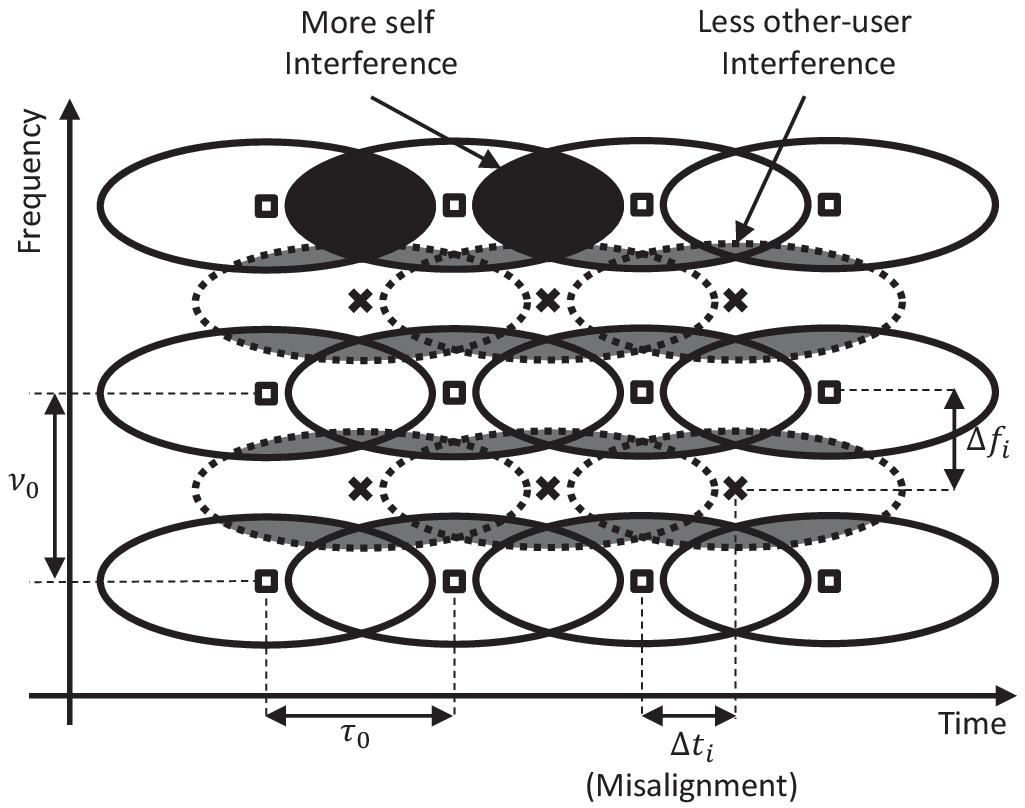}
\label{fig:i2}}
\caption{Illustration for the trade-off between self-interference and other-user interference with the concept of POT. The desired signal and  interfering signal are represented as solid and  dashed lines, respectively. 
}
\label{fig:tradeoff_self_or_other}
\end{figure}
\begin{figure}[!t]
\centering
\subfloat[ Impact of timing misalignment when Gaussian filter is employed.]{\includegraphics[width=3.5in]{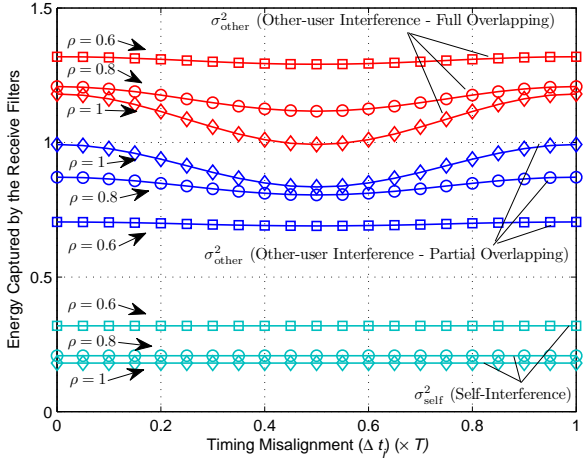}
\label{fig:timingMisalignment_self}}\\
\subfloat[ Trade-off between self-interference and other-user interference.]
{\includegraphics[width=3.5in]{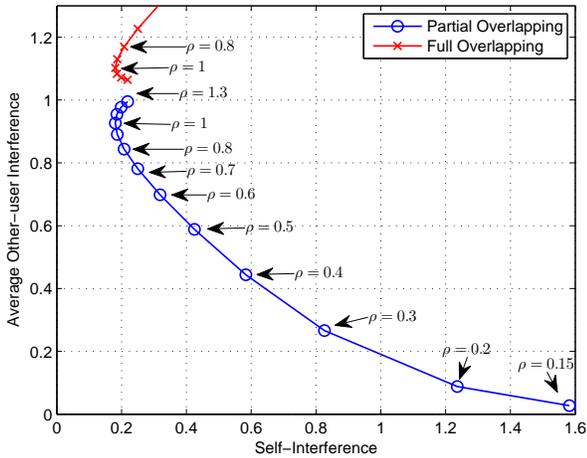}\label{fig:tradeOff_analysis_self}
}
\caption{Other-user interference mitigation without loss in spectral efficiency and power, but at the expense of self-interference.
}
\label{fig:interference_noloss_equalizer}
\end{figure}

In order to understand the utilization of \ac{POT} with non-orthogonal schemes, assume that $\timeSpacingVariable\frequencySpacingVariable=1$ and the transmit pulse shape and the receive filter are Gaussian filters. Gaussian filter is the optimally-concentrated pulse in time-frequency domain and it is expressed as
\begin{align}
\pulseShaping[\timeSymbol]=
(2\gaussianRollOff)^{1/4}e^{-\pi\gaussianRollOff\timeSymbol^2}~,
\end{align}
where $\gaussianRollOff$ is the control parameters for the dispersion of the pulse in time and frequency and $\gaussianRollOff>0$. While the selection of $\gaussianRollOff = 1$ yields a Gaussian filter that has isotropic dispersion in time and frequency, smaller $\gaussianRollOff$ causes more dispersion in time domain and less dispersion in frequency domain. Since Gaussian filter is not a Nyquist filter, consecutive symbols overlap more with smaller $\gaussianRollOff$, yielding more self-interference in time, i.e., \ac{ISI}. 
However, introducing more \ac{ISI} is also beneficial to mitigate the other-user interference,  when \ac{POT} is considered, as illustrated in \figurename~\ref{fig:tradeoff_self_or_other}\subref{fig:i1} and \figurename~\ref{fig:tradeoff_self_or_other}\subref{fig:i2} for $\intentionalCFO=\frequencySpacingVariable/2$. In other words, non-orthogonal schemes yield a trade-off between the other-user interference and self-interference by exploiting the \ac{POT}.

Considering the density of the symbols on time-frequency plane of the victim \ac{RP}, it is important to emphasize the differences between \ac{FTN} signaling \cite{Anderson_2013} and \ac{POT} with non-orthogonal schemes.  In \ac{FTN} signaling, the density of the symbols in time-frequency plane is increased more than Nyquist rate, i.e., $\timeSpacingVariable\frequencySpacingVariable<1$, intentionally. However,  each individual link operates at the Nyquist rate, i.e., $\timeSpacingVariable\frequencySpacingVariable=1$, for \ac{POT}.
The time-frequency plane of the victim \ac{RP} is packed due to the aggressors' signals, which is common in co-channel interference problems. In addition, \ac{POT} does not suggest a structured symbol packing into the time-frequency plane, as in \ac{FTN} signaling. It allows timing misalignment among the individual links. 

Similar to the investigations given in Section \ref{subsec:tradeoff1}, $\intentionalCFO$ is set to $\frequencySpacingVariable/2$ and  $\timingMisalignmentTime$ is swept for one symbol period.
Impact of timing misalignment is given in \figurename~\ref{fig:interference_noloss_equalizer}\subref{fig:timingMisalignment_self}.
In case of full overlapping, when $\timingMisalignmentTime=0$, the receive filter has full correlation with the concentric symbol of the aggressor and partial correlations with the neighboring symbols. Hence, the total energy after the correlation becomes more than 1. In case of partial overlapping, receive filters only capture energy from only the neighboring symbols of aggressors, which yields that $\meanSelfInterferenceGain<1$ as in \figurename~\ref{fig:interference_noloss_equalizer}\subref{fig:timingMisalignment_self}.  In \figurename~\ref{fig:interference_noloss_equalizer}\subref{fig:tradeOff_analysis_self}, the trade-off between self-interference and average other-user interference is given, assuming uniform timing misalignment. As in \figurename~\ref{fig:interference_noloss_equalizer}\subref{fig:tradeOff_analysis_self}, Gaussian filter provides flexible trade-off between self-interference and other-user interference.

There are two potential drawbacks of this approach: 1) necessity for a self-interference cancellation method, e.g., equalization, since the filters do not satisfy Nyquist criterion and 2) colored noise  due to the non-orthogonal receiver filters.
For the first issue, the introduced complexity due to self-interference cancellation method might be preferable in comparison with the complexities of the methods for handling {\em asynchronous} other-user interference.
For the second point, note that non-orthogonal transformations always introduce correlation between samples \cite{Kozek_1998}. If a sequence-based equalizer, e.g., \ac{MLSE}, is employed, a whitening filter should also be utilized to improve the  performance of the receiver. Note that assuming the small link distances for the pairs, noise might become a secondary problem when interference is a dominant issue.

\section{Average BER Analysis}
\label{sec:BERana}
In this section, average \ac{BER} analysis is provided for \ac{POT} for orthogonal schemes that do not introduce self-interference as discussed in Section \ref{subsec:tradeoff1}. To obtain theoretical (but tractable) \ac{BER} analysis, a useful method for \ac{BER} calculations introduced in \cite{hamdi_2007} is combined with spatial \ac{PPP} approaches \cite{Win_2009,Pinto_2010,andrews_2011}. First, \ac{BER} is expressed along \ac{SINR} given in \eqref{Eq:SINR}. Then, its expected value is obtained considering other-user interference. Its computation complexity is significantly reduced by using spatial \ac{PPP} and ambiguity function. For the trade-off introduced in Section \ref{subsec:tradeoff2}, investigation on \ac{BER} performance is performed through the numerical analysis in Section \ref{sec:numericalResults}, since achievable \ac{BER} performance  depends highly on the employed self-interference cancellation method at the receiver.

Closed-form expression for \ac{BER} of a square $\modulationOrder$-\ac{QAM} in \ac{AWGN} channel is readily available in the literature and it is given by
\begin{align}
\berOperator[SNR]&={\sum_{\constellationIndex}^{\sqrt{\modulationOrder}-2} \modulationConstant \erfcOperator[{(2\constellationIndex+1)\sqrt{\frac{SNR}{2}}}]}
\label{Eq:AWGNBER} 
\end{align}
where $\modulationOrder$ is the constellation size, $\modulationConstant$ are the constants depending on the modulation order and $\sum_{\constellationIndex=0}^{\sqrt{\modulationOrder}-2} \modulationConstant = 1/2$ \cite{Kyongkuk_2002}. For instance, $\modulationConstant=\{1/2\}$ and $\constellationIndex=\{0\}$ for $4$-\ac{QAM} and  
$\modulationConstant=\{3/8, 2/8, -1/8\}$ and $\constellationIndex=\{0,1,2\}$  for $16$-\ac{QAM}, respectively.

By substituting \eqref{Eq:SINR} into \eqref{Eq:AWGNBER}, \ac{BER}  is  obtained  for given $\interferenceFunction$, $\interferenceGain[\epsilon]$, and $\victimDistanceOwnUser$ as 
\begin{align}
&\berOperator[{\EbN0|\interferenceGain[\epsilon], \interferenceFunction,\victimDistanceOwnUser}]\nonumber\\&= \displaystyle\sum_{\constellationIndex=0}^{\sqrt{\modulationOrder}-2} \modulationConstant\erfcOperator[\frac{2\constellationIndex-1}{\sqrt{2}}\sqrt{\frac{\interferenceGain[\epsilon]}{\displaystyle
\interferenceFunction+\frac{\modulationOrder-1}{3\log_2{\modulationOrder}}\frac{1}{\EbN0}}
}]~.
\label{Eq:BERint}
\end{align}
Since the target is to calculate average \ac{BER} under interference, the terms, $\interferenceFunction$, and $\interferenceGain[\epsilon]$, have to be averaged out.
In order to obtain average \ac{BER}, we refer to following lemma introduced in \cite{hamdi_2007}:

\smallskip
{\em Lemma-I: Let $x$ and $y$ be unit-mean exponential and arbitrary non-negative random variables, respectively. Then
\begin{align}
\expectedOperator[{\erfcOperator[{\sqrt{\frac{x}{ay+b}}}]}][x,y] = 1-\frac{1}{\sqrt{\pi}}\int_0^{\infty} \frac{e^{-z(1+b)}}{\sqrt{z}}\LaplaceOperator[y][az] \integrald z \nonumber
\end{align}
where
$\LaplaceOperator[y][z] = \expectedOperator[e^{-yz}][y]$ is the \ac{MGF} with negative argument (or Laplace transformation) of random variable $y$. }

\smallskip

If {\em Lemma-I} is applied to \eqref{Eq:BERint} (see e.g., \cite{34_Medjahdi_TWC_2011,Hamdi_2009,hamdi_2007,hamdi_2010}), average \ac{BER} is obtained as 
\begin{align}
&\berOperator[{\EbN0|,\victimDistanceOwnUser}]\nonumber\\&= \displaystyle\sum_{\constellationIndex=0}^{\sqrt{\modulationOrder}-2} \modulationConstant\left(
1-\frac{1}{\sqrt{\pi}}\int_0^{\infty} \frac{e^{-z(1+\frac{2}{(2\constellationIndex-1)^2}\frac{\modulationOrder-1}{3\log_2{\modulationOrder}}\frac{1}{\EbN0})}}{\sqrt{z}} \nonumber\right.\\&~~~~~~~~~~~~~~~~~~~~~~~~~~~~~~~\left.\times \LaplaceOperator[\interferenceFunction][{\frac{2z}{(2\constellationIndex-1)^2}}] \integrald z\right)~,
\label{Eq:averageBERint_step1}
\\
&=\displaystyle\frac{1}{2}-\frac{1}{\sqrt{\pi}}\sum_{\constellationIndex=0}^{\sqrt{\modulationOrder}-2}\modulationConstant\int_0^{\infty} \frac{e^{-z(1+\frac{2}{(2\constellationIndex+1)^2}\frac{\modulationOrder-1}{3\log_2{\modulationOrder}}\frac{1}{\EbN0})}}{\sqrt{z}}
\nonumber\\&~~~~~~~~~~~~~~~~~~~~~~~~~~~~~~\times
\LaplaceOperator[\interferenceFunction][\frac{2z}{(2\constellationIndex+1)^2}] \integrald z~.
\label{Eq:averageBERint}
\end{align}
Therefore, the complexity introduced by \eqref{Eq:BERint} reduces to calculate Laplace transformation of $\interferenceFunction$. In the following subsections, Laplace transformation of $\interferenceFunction$ is calculated in cases of single aggressor and multiple aggressors.

\subsection{Single Aggressor}
If only $\aggressorindex$th aggressor is considered, the Laplace transformation of the total interference is obtained as
\begin{align}
\LaplaceOperator[\interferenceFunction][z] &= \expectedOperator[e^{\textrm{-}z{\interferenceFunction}}][{\interferenceFunction}] \nonumber\\
&\stackrel{(a)}{=}\expectedOperator[e^{\textrm{-}z{\interferenceFunctionSelf}}][{\interferenceFunctionSelf}] \times
\expectedOperator[e^{\textrm{-}z{\interferenceFunctionAggressor}}][\interferenceFunctionAggressor]=\expectedOperator[e^{\textrm{-}z{\interferenceFunctionAggressor}}][\interferenceFunctionAggressor]\nonumber\\
&\stackrel{(b)}{=}
\int_{0}^{\timeSpacingVariable}
\frac{ \pdfOperator[\timingMisalignmentTime] }{
{1+z
{\victimDistanceOwnUser}^{\frac{\aggressorPathlossSlope-\pathlosscompensationParameter\aggressorPathlossSlope}{10}}
\aggressordistanceOwnUser^{\frac{\pathlosscompensationParameter\aggressorPathlossSlope}{10}}{\aggressordistance[]}^{\frac{-\aggressorPathlossSlope}{10}}
\meanInterferenceGain[{\aggressorindex}][{\timingMisalignmentTime,\intentionalCFO}] }}\integrald\timingMisalignmentTime
\label{Eq:singleAggressorP}
\end{align}
where (a) follows from the independent assumption of random variables $\interferenceFunctionAggressor$ and $\interferenceFunctionSelf$ and the assumptions of zero self-interference via orthogonal schemes, (b) is because of the exponential distribution of $\interferenceFunctionAggressorAll$ and the randomness of timing misalignment. Considering the uniform timing misalignment assumption and being a constant function of $\meanInterferenceGain[{\aggressorindex}][{\timingMisalignmentTime,\intentionalCFO}] $ respect to  $\timingMisalignmentTime$, as  in \figurename~\ref{fig:interference_lossy}\subref{fig:fmt}, \eqref{Eq:singleAggressorP} is simplified as
\begin{align}
\LaplaceOperator[\interferenceFunction][z] =
\frac{ 1 }{
{1+z
{\victimDistanceOwnUser}^{\frac{\aggressorPathlossSlope-\pathlosscompensationParameter\aggressorPathlossSlope}{10}}
\aggressordistanceOwnUser^{\frac{\pathlosscompensationParameter\aggressorPathlossSlope}{10}}{\aggressordistance[]}^{\frac{-\aggressorPathlossSlope}{10}}
\meanInterferenceGain[{\aggressorindex}][{\intentionalCFO}] }}
\label{Eq:singleAggressor}
\end{align}

\subsection{Multiple Aggressors}
When multiple aggressors exist in the network, the choice of $\intentionalCFO$  within the link affects the performance of \ac{POT}. 
In order to avoid the coordination, it is assumed that $\intentionalCFO$ is selected randomly from the set $\iCFOset$ given by  $[\iCFO[0], \iCFO[1], \dots, \iCFO[\iCFOindex],\dots]$. The selection is performed based on a \ac{PMF} where $\iCFOprobability[\iCFOindex]$ corresponds to the probability of $\iCFOindex$th intentional \ac{CFO}. Based on this assumption, the Laplace transformation of the total interference is obtained as
\begin{align}
\LaplaceOperator[\interferenceFunction][z] &= \expectedOperator[e^{\textrm{-}z{\interferenceFunction}}][{\interferenceFunction}] \nonumber\\
&\stackrel{(a)}{=}\expectedOperator[e^{\textrm{-}z{\interferenceFunctionSelf}}][{\interferenceFunctionSelf}] \times
\expectedOperator[e^{\textrm{-}z{\sum_{\aggressorindex\in\ppp }\interferenceFunctionAggressor}}][\ppp,\interferenceFunctionAggressor]\nonumber\\
&\stackrel{(b)}{=}
\expectedOperator[e^{\textrm{-}z{\sum_{\aggressorindex\in\ppp }\interferenceFunctionAggressor}}][\ppp,\interferenceFunctionAggressor]
=
\expectedOperator[\prod_{\aggressorindex\in\ppp} {\expectedOperator[e^{\textrm{-}z{\interferenceFunctionAggressor}}][\interferenceFunctionAggressor]}][\ppp] \label{Eq:MGFmMS}\\
&\stackrel{(c)}{=} 
\exp\left[- 2\pi\aggressorDensity \int_{\aggressorMinimumDistance}^{\infty}
\left(
1-{\expectedOperator[e^{\textrm{-}z{\interferenceFunctionAggressor}}][\interferenceFunctionAggressor]} 
\right) \aggressorDistanceVariable
\integrald\aggressorDistanceVariable
\right]
\label{Eq:MGFmMSe}
\end{align}
where (a) follows from the independent assumption of random variables $\interferenceFunctionAggressorAll$ and $\interferenceFunctionSelf$, (b) is because of zero self-interference via orthogonal schemes, and (c) is caused by the probability generating functional of \ac{PPP}, which states $\expectedOperator[{\prod_{\aggressorindex\in\ppp} f(x)}][\ppp] = \exp{\int_{\mathbb{R}^2}(1-f(x))\integrald x}$ for an arbitrary function $f(x)$ and the assumption of  \ac{iid} interference from each aggressor $\interferenceFunctionAggressor$ and independent $\ppp$ from other random variables in the interference function $\interferenceFunctionAggressorAll$\cite{andrews_2011}.
Considering randomness of aggressors' distances $\aggressordistanceOwnUser$, $\expectedOperator[e^{\textrm{-}z{\interferenceFunctionAggressor}}][\interferenceFunctionAggressor]$ is obtained as
\begin{align}
\expectedOperator[e^{\textrm{-}z{\interferenceFunctionAggressor}}][\interferenceFunctionAggressor]=&
\expectedOperator[e^{\textrm{-}z{\frac{{\aggressorInterferencePower}  }{{\aggressorRxPower}} \interferenceGain[\aggressorindex]}}][{\aggressordistanceOwnUser, \interferenceGain[\aggressorindex]}]\nonumber\\
=&\sum_{\iCFOindex} \iCFOprobability[\iCFOindex]
\int_0^{\infty} \frac{\pdfOperator[\aggressordistanceOwnUserVariable]
}{
{1+z{\victimDistanceOwnUser}^{\frac{\aggressorPathlossSlope-\pathlosscompensationParameter\aggressorPathlossSlope}{10}}
\aggressordistanceOwnUserVariable^{\frac{\pathlosscompensationParameter\aggressorPathlossSlope}{10}}{\aggressorDistanceVariable}^{\frac{-\aggressorPathlossSlope}{10}} \meanInterferenceGain[{\aggressorindex}][{\iCFO[\iCFOindex]}]}}\integrald \aggressordistanceOwnUserVariable
\label{Eq:MGFmMSuplink}
\end{align}
which is based on the Laplace transformation of an exponentially disturbed random variable, uniform timing misalignment assumption, and being a constant function of $\meanInterferenceGain[{\aggressorindex}][{\timingMisalignmentTime,\intentionalCFO}] $ respect to  $\timingMisalignmentTime$. 
In \eqref{Eq:MGFmMSuplink}, the \ac{PDF} of $\aggressordistanceOwnUser$ is given by $\pdfOperator[\aggressordistanceOwnUserVariable]=2\pi\aggressorDensity\aggressordistanceOwnUserVariable e^{-\aggressorDensity\pi\aggressordistanceOwnUserVariable^2} $ \cite{novlan_2012}. Then,  $\LaplaceOperator[\interferenceFunction][z]$ is obtained as 
\begin{align}
&\LaplaceOperator[\interferenceFunction][z]=
\displaystyle
\exp\left[- 2\pi\aggressorDensity \int_{\aggressorMinimumDistance}^{\infty}\nonumber \right.\\ & \left.
\left(  
1-\sum_{\iCFOindex} \iCFOprobability[\iCFOindex]
\int_0^{\infty} \frac{
2\pi\aggressorDensity\aggressordistanceOwnUserVariable e^{-\aggressorDensity\pi\aggressordistanceOwnUserVariable^2}
}{
{1+z{\victimDistanceOwnUser}^{\frac{\aggressorPathlossSlope-\pathlosscompensationParameter\aggressorPathlossSlope}{10}}
\aggressordistanceOwnUserVariable^{\frac{\pathlosscompensationParameter\aggressorPathlossSlope}{10}}{\aggressorDistanceVariable}^{\frac{-\aggressorPathlossSlope}{10}} \meanInterferenceGain[{\aggressorindex}][{\iCFO[\iCFOindex]}]}}\integrald \aggressordistanceOwnUserVariable
\right) \aggressorDistanceVariable
\integrald\aggressorDistanceVariable
\right]
\label{Eq:MGFuplinkdownlink}
\end{align}
by substituting \eqref{Eq:MGFmMSuplink} into \eqref{Eq:MGFmMSe}. 
Note that \eqref{Eq:MGFuplinkdownlink} does not always yield a closed-form solution since $\int_0^\infty \frac{xe^{-ax^2}}{1+b x^c} \integrald x$  produces an expression in terms of standard mathematical functions depending on $a,b$, and $c$. Nonetheless, \eqref{Eq:MGFuplinkdownlink} does not require Monte Carlo simulations.

\section{Numerical Results}

\label{sec:numericalResults}
Numerical results are given in order to validate analytical findings with simulations and to investigate the performance of uncoordinated networks along with \ac{POT}. In the simulations, \ac{POT} with orthogonal schemes and \ac{POT} with non-orthogonal schemes are exhibited by utilizing \ac{FMT} with \ac{RRC} filter and zero forcing equalization and by using \ac{NOFDM} with Gaussian filter and symbol-spaced \ac{MLSE} equalization, respectively. For \ac{MLSE}, 7 taps are utilized for each subcarrier and trace-back depth for \ac{MLSE} is set to 20.  Unless otherwise stated, numerical result are obtained for Rayleigh channels.

\begin{figure}[!t]
\centering
\subfloat[ RRC/FMT/4QAM (Solid lines: Analytical results based on \eqref{Eq:averageBERint} and \eqref{Eq:singleAggressor}).]{\includegraphics[width=3.5in]{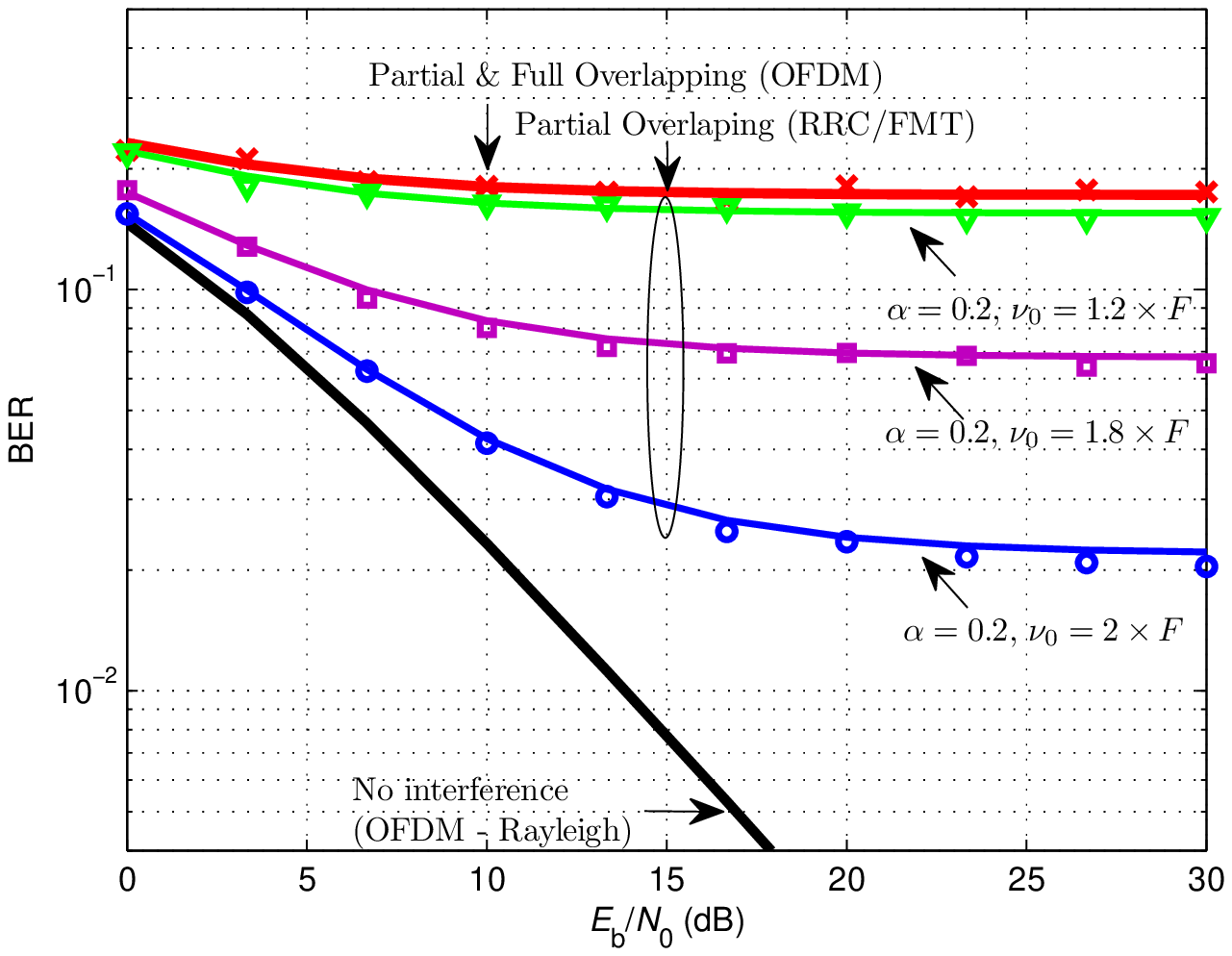}
\label{fig:ofdm_single}}\\
\vspace{\figureDistance}
\subfloat[ Gaussian/NOFDM/4QAM. ]{\includegraphics[width=3.5in]{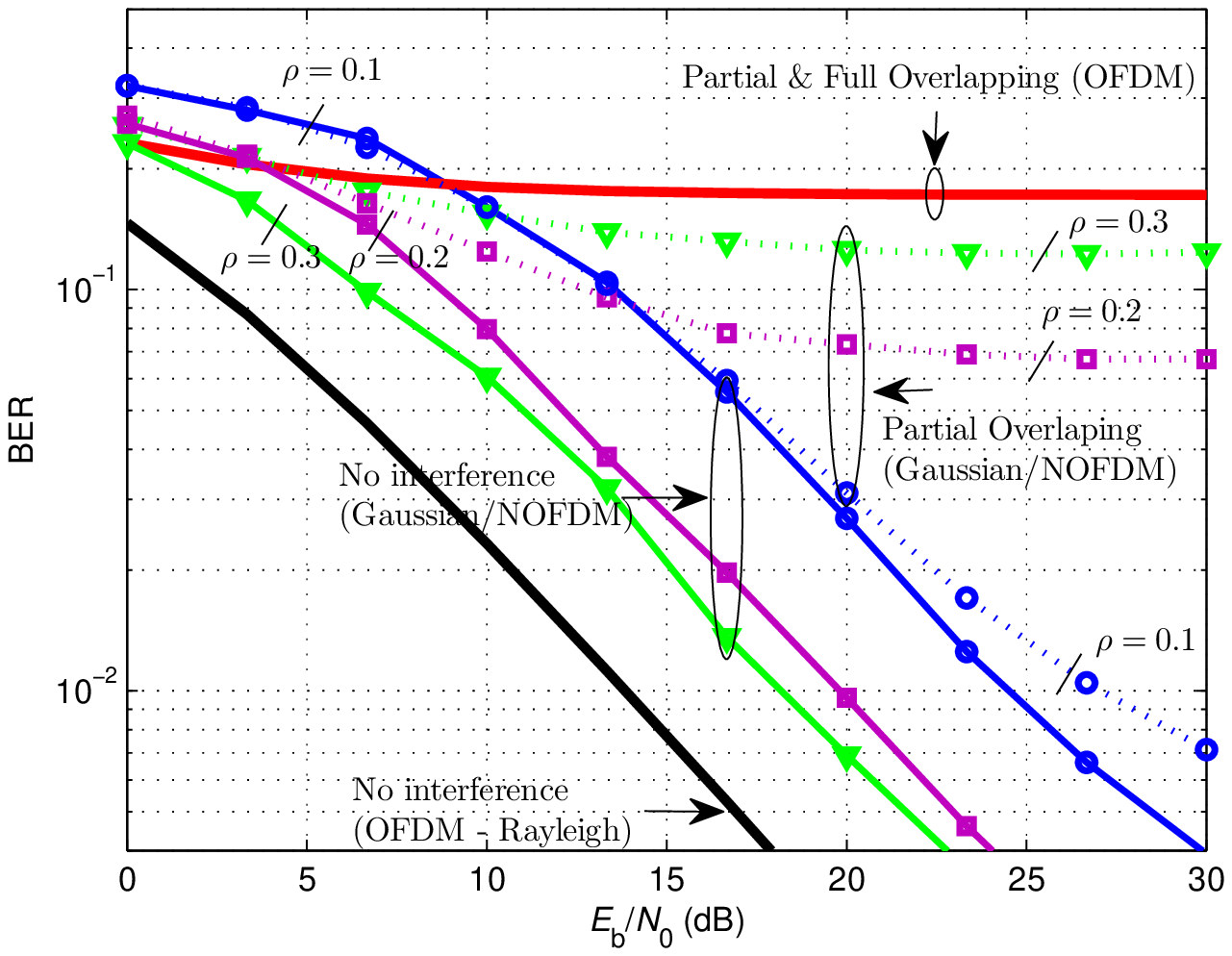}
\label{fig:nofdm_single}}
\caption{BER performance with partial overlapping when there is a single aggressor.}
\label{fig:ber_single_aggressor}
\end{figure}
\begin{figure}[!t]
\centering
\subfloat[ RRC/FMT/4QAM (Solid lines: Analytical results based on \eqref{Eq:averageBERint} and \eqref{Eq:MGFuplinkdownlink}).]{\includegraphics[width=3.5in]{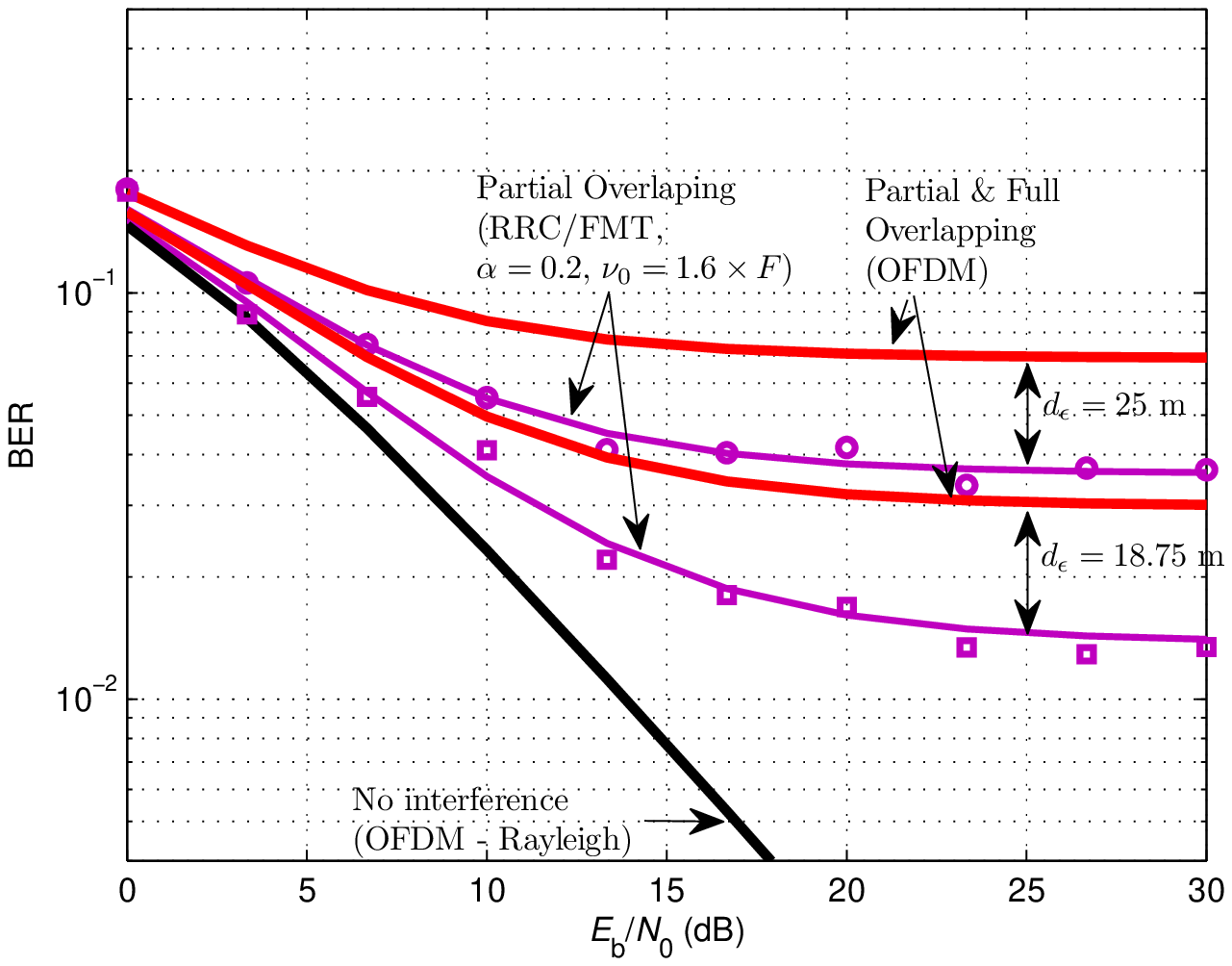}
\label{fig:ofdm_multi}}\\
\vspace{\figureDistance}

\subfloat[ Gaussian/NOFDM/4QAM.]{\includegraphics[width=3.5in]{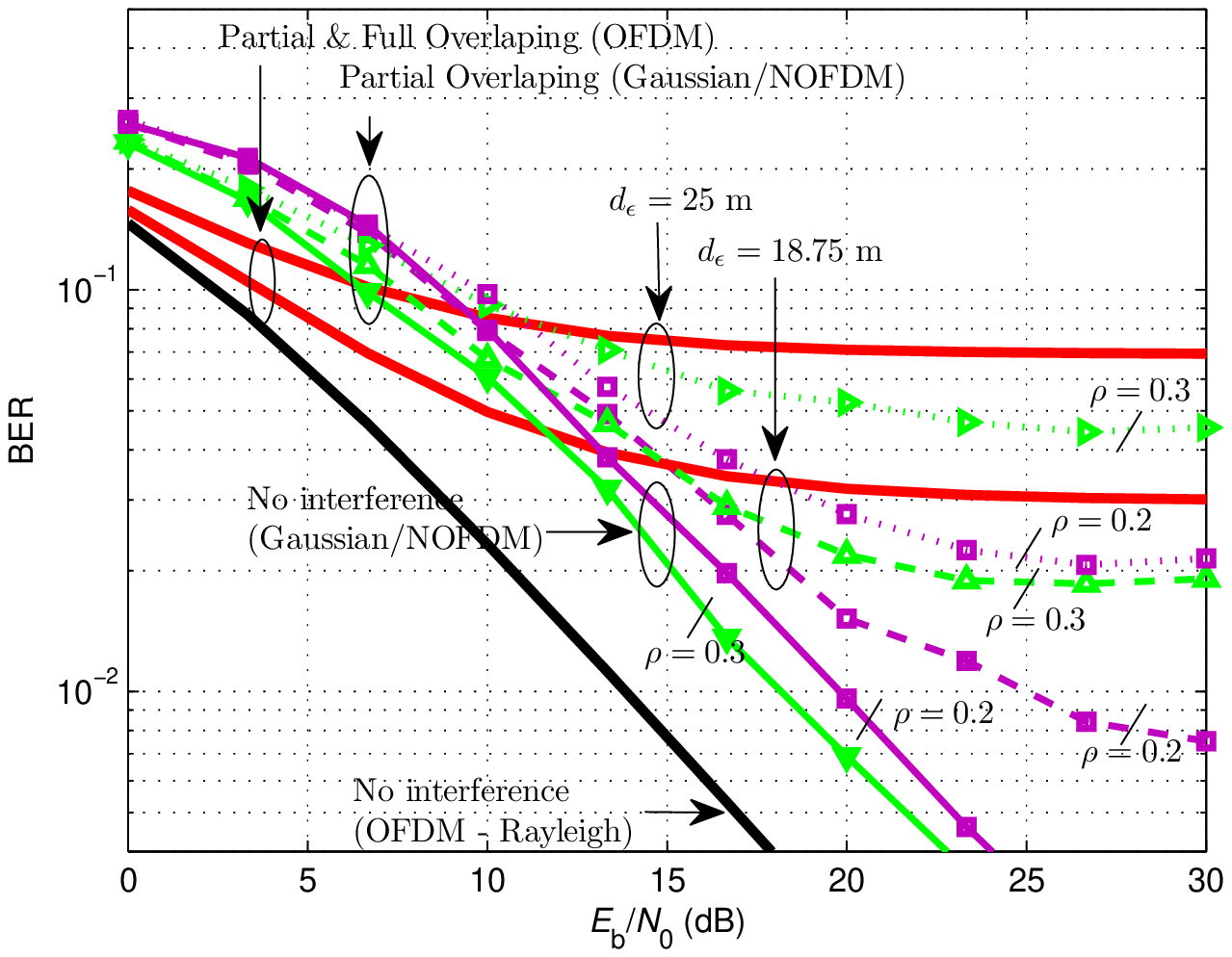}
\label{fig:nofdm_multi}}
\caption{BER performance with partial overlapping when there are multiple aggressors modeled with \ac{PPP}. }
\label{fig:ber_multi_aggressor}
\end{figure}

In \figurename~\ref{fig:ber_single_aggressor}, impact of partial overlapping is presented in Rayleigh channel for the aforementioned trade-offs when a dominant aggressor interrupts the transmission with the equal received signal power (i.e., \ac{SIR} is set to $0$~dB). In \figurename~\ref{fig:ber_single_aggressor}\subref{fig:ofdm_single},  $\rollOff$ is set to $0.2$ and the subcarrier spacing is swept from $1.2\times\frequencySpacing$ to $2\times\frequencySpacing$, referring to the \ac{POT} with orthogonal schemes. 
Also, simulation results are verified with the theoretical results based on \eqref{Eq:averageBERint} and \eqref{Eq:singleAggressor}. 
As it can be seen in \figurename~\ref{fig:ber_single_aggressor}\subref{fig:ofdm_single}, efficacy of \ac{POT} in the \ac{BER} performance increases with the subcarrier spacing, which also causes less spectrally efficient schemes.  In \figurename~\ref{fig:ber_single_aggressor}\subref{fig:nofdm_single}, the same analysis is performed for \ac{NOFDM} to address the \ac{POT} with non-orthogonal schemes.  When other-user interference do not exist, orthogonal schemes reach the Rayleigh bound and introduce superior \ac{BER} performance compared to non-orthogonal schemes. This is mainly because of the fact that \ac{MLSE} loses its optimality under the colored noise caused by the non-orthogonal  transformation at the receiver. However, when the other-user interference exists, orthogonal schemes capture the total amount of the other-user interference and \ac{BER} performance deteriorates significantly. In contrast to orthogonal waveforms, non-orthogonal schemes become notable with the concept of \ac{POT} under the other-user interference. By providing sufficient non-orthogonality, e.g., $\gaussianRollOff = 0.1$, \ac{BER} performance remains the same of the case without other-user interference for \ac{NOFDM} for low to medium \ac{SNR}, as it can be seen in \figurename~\ref{fig:ber_single_aggressor}\subref{fig:nofdm_single}. Essentially, the results show that \ac{BER} performance is enhanced without sacrificing the spectral efficiency at the expense of complexity at the receiver.

In \figurename~\ref{fig:ber_multi_aggressor}, impact of \ac{POT} on \ac{BER} performance is shown when there are multiple aggressors. In the simulation,
the path loss is modeled with the parameters given in \cite{Report_ITU-R_M.2135} as
\begin{align}
L(d)=&11.8+ 45\log_{10}(\carrierFrequency)+40\log_{10}(d/1000)
\label{Eq:pathloss}
\end{align}
where $\carrierFrequency$ is the carrier frequency in MHz (3500 MHz) and $d$ is the distance in meters. Using given parameters, the path loss formula is calculated as $L(\cdot) = 51.3 + 40 \log (\cdot)$ where the argument is in terms of meters. Accordingly, $\aggressorPathlossConstant$ and $\aggressorPathlossSlope$ are set to $51.3$ and $40$, respectively.  The intensity of \ac{TP} and $\aggressorMinimumDistance$ are set to $1/(\pi 50^2)$ and $25$~m, respectively. In order to see the best possible \ac{BER} performance, all aggressors' signals are partially overlapped with the desired signal. Then, \ac{BER} curves are obtained  for different victim link distance $\victimDistanceOwnUser$.  
As expected, \ac{BER} is directly related to the user distance. Especially, the degradation becomes severe for the users located at far distances. In \figurename~\ref{fig:ber_multi_aggressor}\subref{fig:ofdm_multi}, it is shown that orthogonal schemes allow better \ac{BER} performance with the concept of \ac{POT} by losing their spectral efficiencies.
Also, simulation results match with the theoretical results based on \eqref{Eq:averageBERint} and \eqref{Eq:MGFuplinkdownlink}.
 In \figurename~\ref{fig:ber_multi_aggressor}\subref{fig:nofdm_multi}, the impact of non-orthogonal schemes on \ac{BER} performance are shown for the same scenario and better \ac{BER} performance is obtained for high $\EbN0$ without any spectral efficiency loss, but complexity at the receiver. Considering \figurename~\ref{fig:ber_single_aggressor}\subref{fig:nofdm_single} and \figurename~\ref{fig:ber_multi_aggressor}\subref{fig:nofdm_multi}, it is important to emphasize that one may obtain the optimum $\gaussianRollOff$, considering the amount of the attainable self-interference and the amount of mitigated other-user interference. Although the selection of $\gaussianRollOff =0.1$ significantly improve the \ac{BER} performance when the amount of the other-user interference is equal to signal power, the same scheme might not yield optimum \ac{BER} performance when other-user interference becomes weaker due to the path loss. Essentially, this issue indicates that there is a point where non-orthogonality starts to be harmful. Therefore, the best selection of $\rho$ depends on the equalizer performance and the amount of the other-user interference.

\section{Concluding Remarks}
\label{sec:conclusion}

In this study, by allowing intentional \ac{CFO} between the interfering links, other-user interference is mitigated in an uncoordinated network without any timing constraints via orthogonal or non-orthogonal  schemes.
For a well-coordinated network, transmission over orthogonal schemes might lead to better performance compared to non-orthogonal schemes due to the absence of self-interference. However, when other-user interference is inevitable and significant in an uncoordinated network, spectral efficiency  has to be sacrificed for orthogonal schemes in order to allow other-user interference mitigation. Specifically, schemes which allow non-overlapping subcarriers in frequency, e.g., \ac{FMT}, complies with the intentional \ac{CFO} approach to avoid timing misalignment problems with \ac{POT}. As opposed to orthogonal waveforms, non-orthogonal schemes come into the prominence along with \ac{POT} for an interesting reason; self-interference problem is easier than other-user interference problem in an uncoordinated networks. By utilizing non-orthogonal waveforms, \ac{POT} is able to change the type of interference from other-user interference to self-interference. This is beneficial when the receiver has proper self-interference cancellation mechanisms. 
Especially, it is promising when two pairs sharing the same spectrum are close to each other.

 Throughout the study, \ac{POT} is presented for two intentional \ac{CFO} levels, i.e., $\fc$ and $\fc+\frequencySpacingVariable/2$. Although the \ac{POT} with two intentional \ac{CFO} levels heuristically matches to two-users scenarios, it might be a suboptimum solution for the multiple-user scenarios. However, it is possible to utilize multiple \ac{CFO} levels
to extend \ac{POT} to multiple-user scenarios. In addition, when the difference between the power levels of interfering signal and desired signal are significantly large, well-known interference cancellation methods, e.g. \ac{SIC}, might provide better results than \ac{POT}. However, the combination of \ac{POT} and interference cancellation techniques can increase the performance substantially. Since \ac{POT} is able to increase the difference between the norm of interference and the norm of desired signal power, \ac{POT} is also able to increase the separability of the signals.

\section*{Acknowledgment}
This study has been supported by InterDigital Communications Inc.

\bibliographystyle{IEEEtran}
\bibliography{pot}
\clearpage

\end{document}